\documentclass[12pt, draftclsnofoot, onecolumn]{IEEEtran}
\usepackage{array}
\usepackage{multirow}
\usepackage{amsthm}
\usepackage{amsmath}
\usepackage{amsfonts}
\usepackage{amssymb}
\usepackage{graphicx}
\usepackage{multirow}
\usepackage{setspace}
\usepackage{cite}
\usepackage{balance}
\usepackage{graphicx}
\usepackage{caption}
\usepackage{subcaption}
\usepackage{algorithmic}
\usepackage{algorithm}

\begin{document}
	\title{Design and Analysis of Optimal Threshold Offloading Algorithm for LTE Femtocell/Macrocell Networks}

	\author{Wei-Han Chen, Yi Ren, Chia-Wei Chang,  and Jyh-Cheng Chen, \textit{Fellow, IEEE}
		\thanks{The authors are with the Department of Computer Science, National Chiao Tung University, Hsinchu 300, Taiwan (E-mail: \{whchen, yiren, chwchang, jcc\}@cs.nctu.edu.tw)}
	}
	
	\maketitle
	\begin{abstract}
		LTE femtocells have been commonly deployed by network operators to increase network capacity and offload mobile data traffic from macrocells. A User Equipment (UE) camped on femtocells has the benefits of higher transmission rate and longer battery life due to its proximity to the base stations. However, various user mobility behaviors may incur network signaling overhead and degrade femtocell offloading capability. To efficiently offload mobile data traffic, we propose an optimal Threshold Offloading (TO) algorithm considering the trade-off between network signaling overhead and femtocell offloading capability. In this paper, we develop an analytical model to quantify the trade-off and validate the analysis through extensive simulations. The results show that the TO algorithm can significantly reduce signaling overhead at minor cost of femtocell offloading capability. Moreover, this work offers network operators guidelines to set optimal offloading threshold in accordance with their management policies in a systematical way.
	\end{abstract}
	

	\section{Introduction}
	\label{intro}
	
	\IEEEPARstart{T}{he} Long Term Evolution-Advanced (LTE-A)~\cite{ts36.300} standardized by the 3rd Generation Partnership Project (3GPP) provides high speed wireless data transmission services: up to 100~Mbit/s peak data rates for high mobility access and 1~Gbit/s for low mobility users. LTE-A together with ubiquitous Internet access offered by various mobile devices leads to explosive growth of mobile data traffic in the past few years. According to the 2014 annual report from AT\&T, mobile data traffic on their network increased 100,000\% from 2007 to 2014~\cite{at&t}. A Cisco report states global mobile data traffic has reached 2.5 exabytes per month at the end of 2014. It also forecasts that mobile data traffic will increase almost tenfold between 2014 and 2019~\cite{cisco}. The unprecedented data traffic growth pushes mobile networks to their operational limits and motivates network operators to seek for efficient solutions that can help relieve network congestion.
	
\begin{figure}[tb]
		\graphicspath{ {./Figures/} }
		\centering
		\includegraphics[width=7cm]{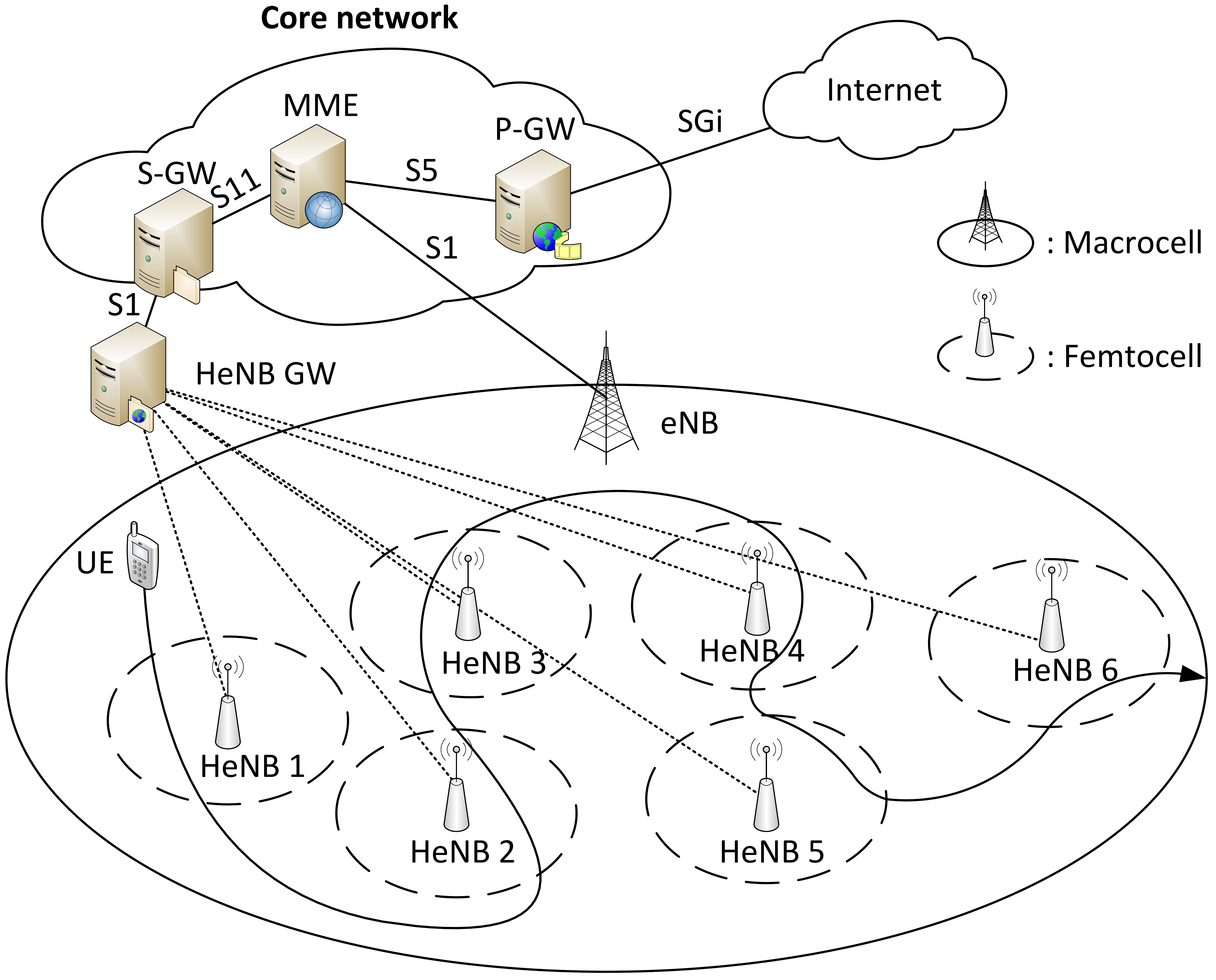}
		\caption{The general LTE-A architecture and an exemplary handover scenario.}
		\label{fig:ExampleScenario}
		\vspace{-3mm}
	\end{figure}

	Data offloading is one of such solutions that alleviate network congestion by moving mobile data traffic from a congested Radio Access Network (RAN) to a capacious target RAN. Generally, data offloading can be classified into two categories: (1) homogeneous network data offloading, and (2) heterogeneous network data offloading. The former offloads mobile data traffic of a User Equipment (UE) within 3GPP RANs, e.g., from a macrocell to a femtocell. In LTE-A network, a macrocell refers to the radio coverage of an evolved NodeB (eNB), and a femtocell (also called a small cell) refers to that of a Home evolved NodeB (HeNB). Fig.~\ref{fig:ExampleScenario}  illustrates the general LTE-A architecture. On the other hand, the latter one offloads the mobile data traffic from a 3GPP RAN to a non-3GPP RAN, e.g., from a macrocell to a Wi-Fi network.
	
	In this paper, we focus on data offloading in homogeneous networks. The emergent femtocell extends the coverage of cellular networks and makes data offloading in homogeneous networks a promising solution. From network operators' point of view, femtocells can extend network coverage, offload mobile data traffic from macrocells, and increase network capacity~\cite{scforum}. From users' point of view, femtocells enable higher transmission rate and longer battery life since UEs receive better signal strength and consume less energy~\cite{chandrasekhar2008femtocell}. With all these attractive benefits, femtocells are commonly accepted by network operators as an effective offloading technique and massively deployed in urban and metropolitan areas. According to~\cite{casestudy} reported in 2014, Korea Telecom (KT) has deployed more than 10,000 small cells in the Seoul metropolitan area and a further 8,000 in greater Seoul, which houses over half population in South Korea.
	
	However, many researchers have pointed out that femtocell offloading may not always guarantee aforementioned benefits and may cause network signaling overhead due to various user mobility behaviors~\cite{chowdhury2009handover, kim2010handover, lee2013reducing, shaohong2009handover, ulvan2010study, zhang2010novel, yang2011handover, whchen2016iscc}.	For instance, if high mobility users reside in a femtocell for a short period of time, only small amount of data can be offloaded. However, they trigger two handover procedures and incur signaling overhead in the core network when they move into and out of a femtocell. For simplicity, we call this behavior as \textit{transient handover}. The transient handovers not only degrade Quality of Experience (QoE) of mobile users but also reduce the availability of femtocells for legitimate users, i.e., reduce system capacity.
	%
	%
	Thus, in order to tackle the transient handover issue and offload mobile data traffic from macrocells in a cost-effective way, it is important to address \textit{when} and \textit{how} should a UE handover into a femtocell when the UE approaches the boundary of the femtocell. It is essentially a trade-off between network signaling overhead and femtocell offloading capability.
	
	In this paper, we propose an optimal algorithm to offload mobile data traffic to femtocells and provide a mathematical model to quantify its performance. Our analytical model and simulation results show consistent findings that the proposed algorithm significantly reduces signaling overhead and sacrifices little femtocell offloading capability. The contributions of this paper are twofold:
	\begin{itemize}
		\item First, we propose an effective algorithm for femtocell traffic offloading considering the trade-off between network signaling overhead and femtocell offloading capability. Moreover, the proposed algorithm is simple. Thus, it is easy to implement.
		\item Second, we provide an analytical model to investigate the performance of our proposed algorithm, which is further validated through extensive simulations. In addition, our mathematical analysis offers guidelines for network operators on setting optimal offloading threshold in a systematic way.
	\end{itemize}
	
	The remaining parts of the paper are organized as follows. Section~\ref{sec:Background} introduces the background. Section~\ref{relatedwork} presents the related work. Section~\ref{proposedscheme} describes the proposed Threshold Offloading (TO) algorithm. The analytical model is presented in Section~\ref{analyticalmodel}.  The simulation and numerical results are given in Section~\ref{simulationandnumericalresults}. Section~\ref{optimaltomechanism} presents the optimal TO algorithm. Section~\ref{conclusion} summarizes the paper.
	
	\section{Background}\label{sec:Background}
	
	Fig.~\ref{fig:ExampleScenario} shows an exemplary scenario that a UE has an active session while passing through the  coverage areas of an eNB and a set of HeNBs, typically with radius within 2,000 meters and 20 meters, respectively~\cite{scforum}. The UE connects to the core network via either the eNB or one of the HeNBs. In the core network, a Mobility Management Entity (MME) handles bearer management, mobility management, and control plane signaling. A Serving Gateway (S-GW) forwards user packets between the UE and the core network. The routing between the core network and the Internet is responsible by a Packet Data Network Gateway (P-GW). Details of the core network, MME, S-GW and P-GW can be found in~\cite{ts36.300}.

\begin{figure}[t]
		\vspace{-3mm}
		\graphicspath{ {./Figures/} }
		\centering
		\includegraphics[width=8cm]{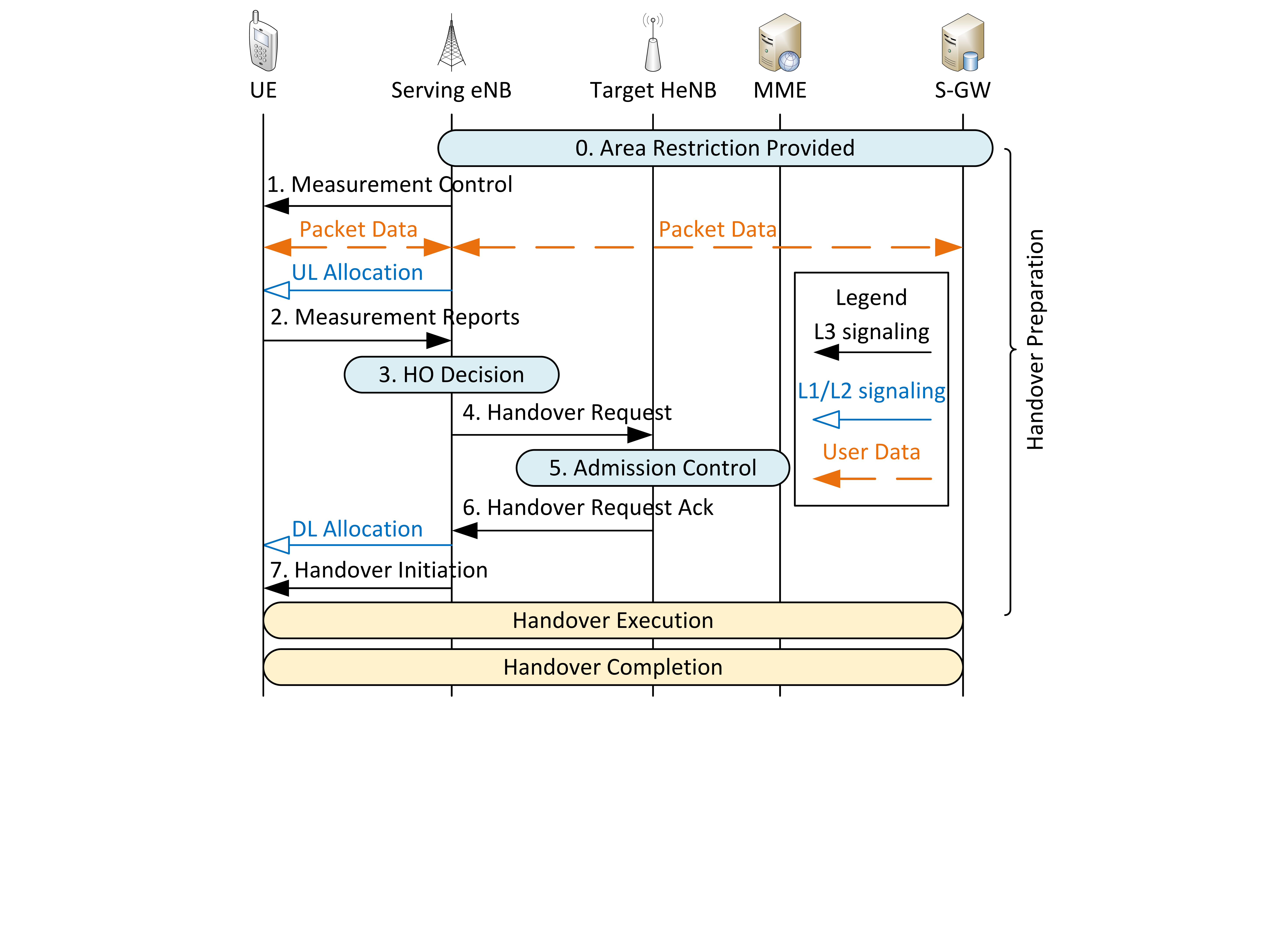}
		\caption{Handover procedure specified in the 3GPP standard.}
		\label{fig:HandoverSpec}
		\vspace{-5mm}
	\end{figure}

	Fig.~\ref{fig:HandoverSpec} shows the handover procedure specified in the 3GPP standard~\cite{ts36.300}, referred to as the baseline handover procedure in this paper. In the figure, solid lines indicate signaling messages, and dashed lines represent user data traffic. There are three phases in the baseline procedure: handover preparation, handover execution, and handover completion. Since the proposed algorithm only involves in handover preparation phase, we elaborate the handover preparation phase as follows:
	
	\begin{itemize}
		\item Step 1: The serving eNB informs the UE in which event the received signal strength should be reported through a configuration message, and the UE keeps track of the received signal strength of its serving cell and neighboring cells.
		\item Step 2: Upon a specified event\footnote{There are several measurement report triggering events defined in the 3GPP standard: event A1-A6, B1, B2, C1, and C2. For example, event A3 is triggered when signal level from neighbor cell becomes amount-of-offset better than serving cell. Further details can be found in~\cite{ts36.331}.} happens, the UE sends measurement reports to the eNB.
		\item Step 3: The serving eNB decides whether to initiate a handover (HO) procedure to the selected target cell based on the information received from the UE measurement reports and the status of the neighboring cells.
		\item Step 4: The serving eNB sends a handover request as well as the UE context (e.g., security and QoS context) to the target HeNB.
		\item Step 5: The admission control ensures that the UE will be served with enough bandwidth and guarantees the UE's QoS requirements. In addition, it checks whether a UE is legitimate to access its resources.
		\item Step 6: If the target HeNB is capable of providing the requested service quality, it informs the serving eNB with a Handover Request Ack.
		\item Step 7: The serving eNB sends a control message to the UE to initiate the handover procedure.
	\end{itemize}
	
	For further understanding of the handover procedure, readers can refer to~\cite{ts36.300} for details.
	
	\section{Related Work}
	\label{relatedwork}
	
	Many studies~\cite{chowdhury2009handover, kim2010handover, lee2013reducing, shaohong2009handover, ulvan2010study, yang2011handover, zhang2010novel,whchen2016iscc} have been proposed for homogeneous network data offloading. The authors in~\cite{chowdhury2009handover} first identified the frequent and unnecessary handover problem for hierarchical network environment. They proposed a Call Admission Control (CAC) mechanism at target HeNB, which significantly reduces the impacts of transient handovers. However, the work did not provide analysis and guidelines for operators on how to set optimal values for the essential parameters. The authors in~\cite{kim2010handover} later extended the problem to hybrid access mode HeNBs and classified users into two groups: pre-registered and unregistered. Unregistered users are not allowed to handover into femtocells so a portion of frequent handovers are efficiently avoided. However, the frequent handover problem still exists for pre-registered users.
	
	In~\cite{zhang2010novel, ulvan2010study, shaohong2009handover}, the authors considered Received Signal Strength (RSS) and UE's moving speed to reduce unnecessary handover executions. The authors in~\cite{shaohong2009handover} designed two novel handover algorithms based on RSS and velocity information from UEs: Velocity and Signal Handover (VSHO) and Unequal Handover (UHO). In VSHO, a UE will handover into a femtocell if its velocity is within the maximum handover velocity level and its RSS level is above the minimum RSS level. As an improvement of VSHO, UHO further considers the difference of signal levels between the eNB and the HeNB. The authors in~\cite{zhang2010novel} took UE mobility state and application type into consideration, where UE mobility behaviors are classified into three states: low (0-15 km/h), medium (15-30 km/h), and high (above 30 km/h). Transient handovers caused by high mobility UEs are avoided effectively. In~\cite{ulvan2010study}, in addition to UE mobility state and application type, the authors further used proactive handover procedure to reduce packet loss for real-time services. However, in LTE networks, velocity measurement of a UE causes extra costs and may not be accurately obtained.

	Recent works~\cite{yang2011handover, lee2013reducing, whchen2016iscc} further considered UE's signal information. In~\cite{yang2011handover}, the authors proposed Double Threshold Algorithm (DTA) which uses two thresholds of Signal to Interference and Noise Ratio (SINR) to determine whether a handover procedure should be performed. Because DTA heavily depends on accurate measurement of SINR, serious interference in densely-deployed metropolitan areas will cause inaccurate measurements and makes it ineffective in those areas. In~\cite{lee2013reducing}, the authors proposed a novel Reducing Handover Cost (RHC) mechanism in that femtocell-to-macrocell handover requests are delayed for a period of time when UEs move out of a femtocell. The work reduces signaling cost caused by transient handovers. However, RHC may not have the desired outcome if the HeNBs are not in close proximity to each other. Our recent work~\cite{whchen2016iscc} proposes a threshold offloading algorithm considering the trade-off between network signaling overhead and femtocell offloading capability. The algorithm significantly reduces signaling overhead at minor cost of femtocell offloading capability. However, the work did not consider how to select an optimal threshold value for the trade-off.
	
	There are many studies on heterogeneous network data offloading, but we only list some representative ones since our focus is on homogeneous network data offloading. A system called Wiffler proposed in~\cite{balasubramanian2010augmenting} exploits the delay tolerance of content and the contacts with fixed Access Points (APs). Wiffler predicts future encounters with APs and defers data transmission only if the transfer can be finished within the application's tolerable threshold and reduce cellular data usage. 	The work~\cite{Han20141} proposes to use cognitive radio techniques to offload mobile traffic to wireless APs, where the trade-off between optimal mobile traffic offloading and energy consumption was studied. The scheme is proved to work well and can save at least 50\% energy of base stations. The authors in~\cite{lee2013mobile} conducted quantitative studies on the benefit of Wi-Fi offloading with respect to network operators and mobile users. They collected trace data for two and a half weeks from Seoul, Korea, and analyzed Wi-Fi availability. The results reveal that non-delayed Wi-Fi offloading could already relieve a large portion of cellular data traffic, and confirm that delayed Wi-Fi offloading has even more promising potential.

	\section{Proposed Threshold Offloading (TO)}
	\label{proposedscheme}
 	
	The transient handover consists of two parts: macrocell-to-femtocell and femtocell-to-macrocell handover.
	The proposed Threshold Offloading (TO) is employed at the serving eNB to prevent undesired macrocell-to-femtocell transient handovers, and so the following femtocell-to-macrocell handovers are also prevented.
	
\begin{figure}[tb]
		\vspace{-3mm}
		\graphicspath{ {./Figures/} }
		\centering
		\includegraphics[width=8.5cm]{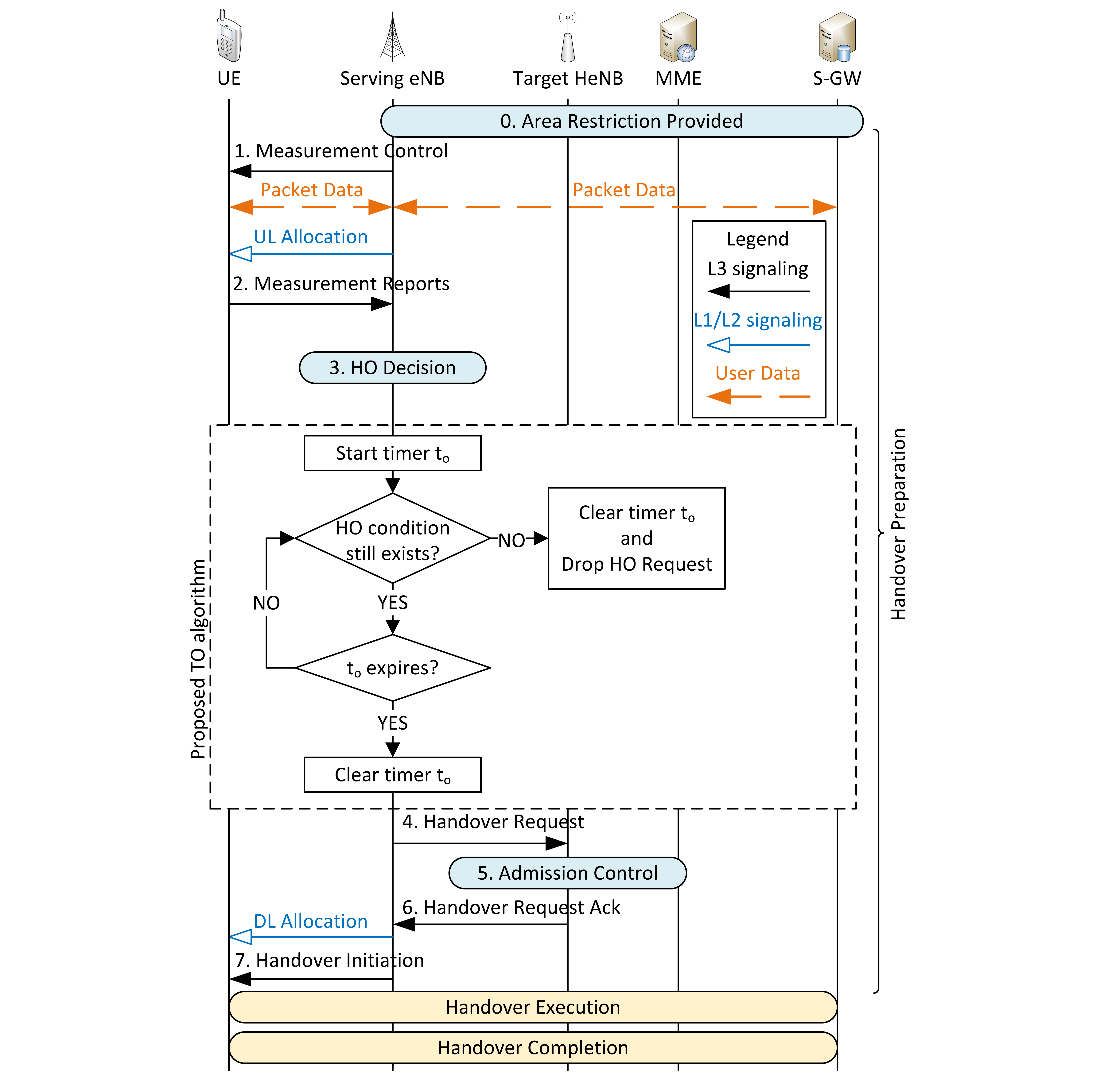}
		\caption{The proposed TO algorithm.}
		\label{fig:HandoverTO}
		\vspace{-5mm}
	\end{figure}

	The proposed TO requires only minor modification of the 3GPP standard, shown in Fig.~\ref{fig:HandoverTO}. When a UE enters a femtocell and meets the handover triggering requirements, the macrocell-to-femtocell handover request is deferred at the serving eNB until a predefined offloading threshold $t_{o}$ is reached. If the UE leaves the femtocell before $t_{o}$ expires, no handover is triggered. Otherwise, the handover request will be sent to the target HeNB if $t_{o}$ has expired and the triggering condition for handover still exists. Fig.~\ref{fig:HandoverTO} illustrates the flow chart for the TO.
	
	The parameter, $t_{o}$, determines the efficiency of the proposed TO. For instance, more signaling overhead can be reduced by setting a larger $t_{o}$, but it will degrade femtocell offloading capability because less time UEs will be offloaded to femtocells. Although the proposed TO sounds intuitive, how to set $t_{o}$ properly is nontrivial. In next section, we present a mathematical model to determine $t_{o}$ in a systematic way.
	
	Note that, in the handover process, $t_{o}$ should be distinguished from the parameter, time-to-trigger (TTT) value, which is mainly for preventing ping-pong effect and radio link failure. Our proposed TO does not use the TTT value to delay handover so that eNBs are able to acquire information of UEs' radio connection in a timely manner. Therefore, radio link failures due to the delayed handover process are avoided.
	
	\section{Analytical Model}
	\label{analyticalmodel}
	
	In this section, we investigate the trade-off between \textit{signaling overhead} and femtocell \textit{offloading capability}. We first propose a mathematical model to quantify the characteristics of the trade-off. Next, the impacts of the offloading threshold $t_{o}$ are studied by two performance metrics: \textit{signaling overhead reduction ratio} $\Theta(t_{o})$ and \textit{offloading capability loss ratio} $\Lambda(t_{o})$.
	
	We define signaling overhead reduction ratio $\Theta(t_{o})$ ($0<\Theta(t_{o})\leq 1$) as:
	\begin{equation}
		\Theta(t_{o}) = \frac{E[N_{b}]- E[N_{t}(t_{o})]}{E[N_{b}]} \ , \label{metric1}
	\end{equation}	
	where $N_{t}(t_{o})$ and $N_{b}$ are the total number of handovers during a session with and without TO, respectively. $\Theta(t_{o})$ indicates at what percentage the proposed TO reduces total number of handovers in a session. The higher $\Theta(t_{o})$ is, the better the TO performs. Next, we define offloading capability loss ratio $\Lambda(t_{o})$ ($0<\Lambda(t_{o})\leq 1$) as:
	\begin{equation}
		\Lambda(t_{o}) = \frac{E[T_{t}(t_{o})]}{E[T_{b}]} \ , \label{metric2}
	\end{equation}	
	where $T_{t}(t_{o})$ and $T_{b}$ are the total time that the session is served in femtocell with and without TO, respectively. No assumptions made on traffic types or traffic distributions, we use time ratio to represent the possibility that a UE's session can be served by femtocells. Since the TO will reduce the amount of time offloaded to femtocell during a session, $\Lambda(t_{o})$ indicates how much offloading capability the TO can remain compared to the baseline scheme in 3GPP standard.
	
	
	There is an inverse relationship between the two performance metrics, so 	
	our design goal is to find an optimal $t_{o}$ such that it can bring $\Theta(t_{o})$ and $\Lambda(t_{o})$ to their possible maxima.
	
	The radio coverage areas of femtocells overlapped with a macrocell may be continuous or discontinuous. The difference between them is that, in discontinuous case, a UE can be at most under one femtocell coverage, therefore there is only one target HeNB to handover. While in continuous case, a UE can be under multiple femtocell coverage, and there are multiple target HeNBs, which complicates UEs' measurement reporting and handover decision. In this study, we first consider the discontinuous case and leave the continuous one as our future work.
	
\begin{figure*}
			\vspace{-5mm}
			\graphicspath{ {./Figures/} }
			\centering
			
			\begin{subfigure}[tb]{0.7\textwidth} \includegraphics[width=\textwidth]{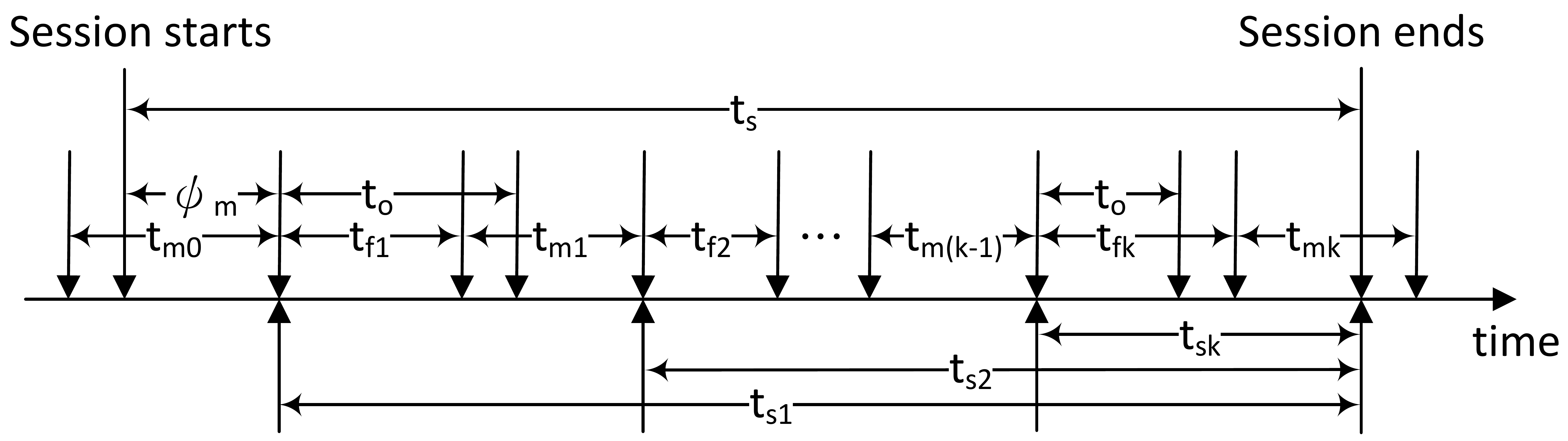} 	
				\caption{Case 1-1: A UE session starts and ends in a macrocell.}
				\label{fig:TimingDiagram_a}
			\end{subfigure} ~ 
			
			\begin{subfigure}[tb]{0.7\textwidth} \includegraphics[width=\textwidth]{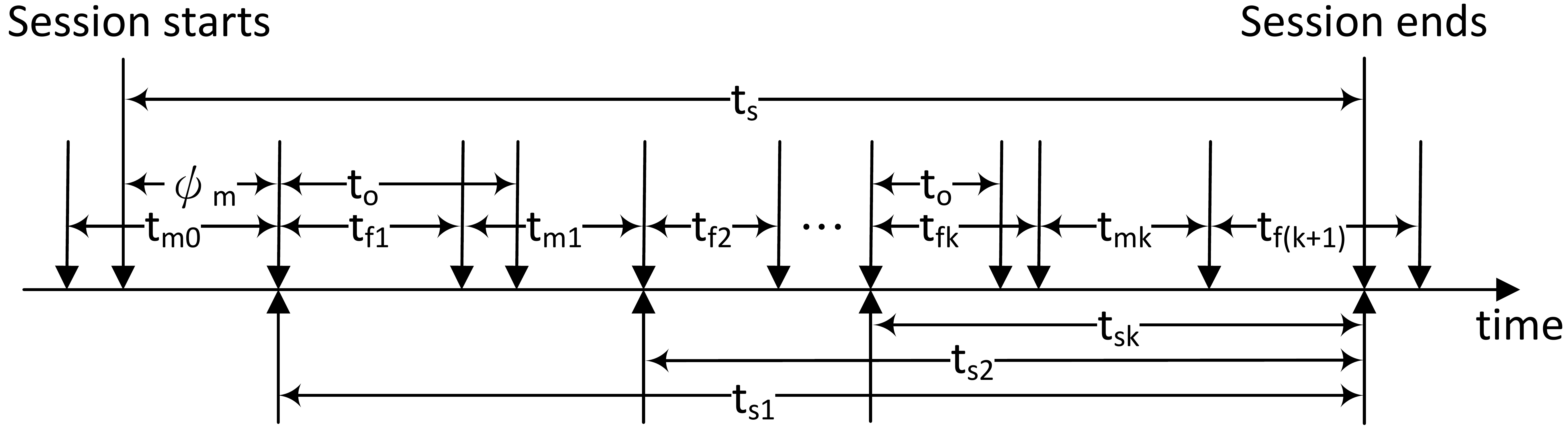}
				\caption{Case 1-2: A UE session starts in a macrocell and ends in a femtocell.}
				\label{fig:TimingDiagram_b}
			\end{subfigure} ~ 
			
			\begin{subfigure}[tb]{0.7\textwidth} \includegraphics[width=\textwidth]{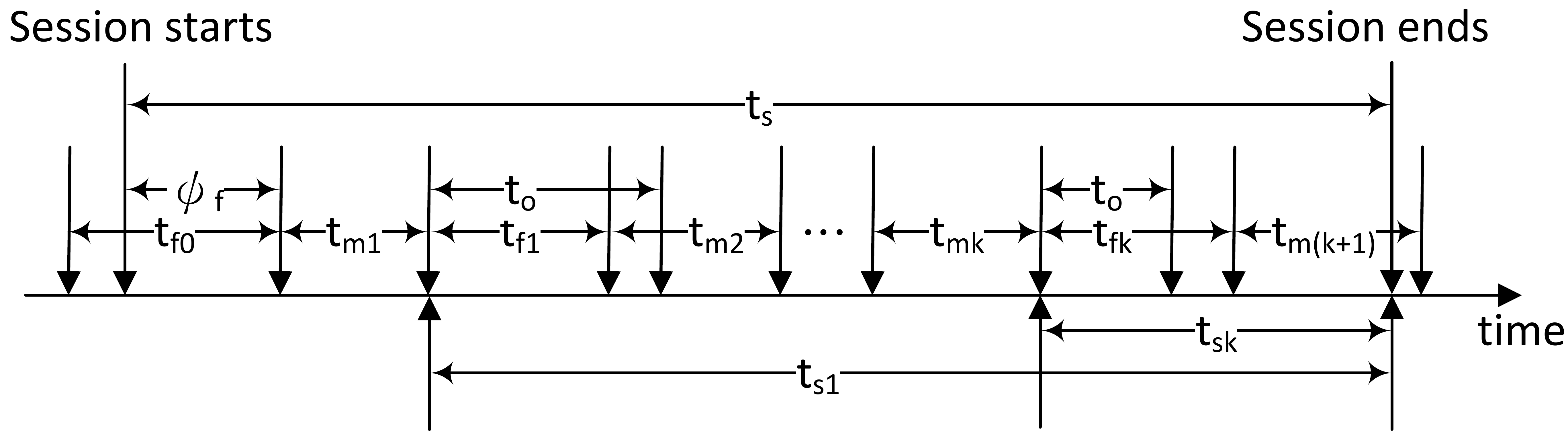}
				\caption{Case 2-1: A UE session starts in a femtocell and ends in a macrocell.}
				\label{fig:TimingDiagram_c}
			\end{subfigure} ~ 
			
			\begin{subfigure}[tb]{0.7\textwidth} \includegraphics[width=\textwidth]{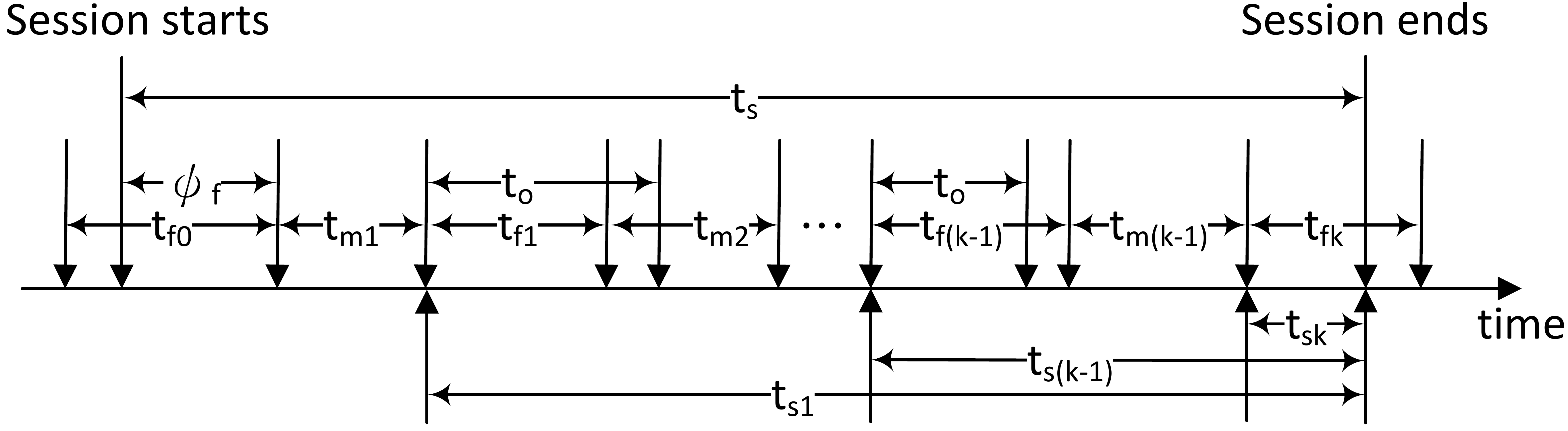}
				\caption{Case 2-2: A UE session starts and ends in a femtocell.}
				\label{fig:TimingDiagram_d}
			\end{subfigure} ~ 
			
			\caption{The timing diagram for the four handover scenarios of a UE session}
			\label{fig:TimingDiagram}
			\vspace{-7mm}
		\end{figure*}

	Because the cell shape (e.g. hexagonal or circular), the cell size, UE moving speed, and their moving direction are hard to be characterized, in this paper, the mobility behavior of a UE is modeled by the length of Cell Residence Time (CRT). This is commonly adopted in previous studies~\cite{lee2013reducing, liou2013investigation}. In our analytical model, when a UE travels in the macrocell, it alternately stays in the macrocell and the femtocells. A UE's session can be categorized into four cases, as shown in Fig.~\ref{fig:TimingDiagram}. In Fig.~\ref{fig:TimingDiagram_a}, a UE's session starts and ends in a macrocell. In Fig.~\ref{fig:TimingDiagram_b}, a UE's session starts in a macrocell but ends in a femtocell. Fig.~\ref{fig:TimingDiagram_c} shows that a UE's session starts in a femtocell and ends in a macrocell. A UE's session starts and ends in a femtocell is depicted in Fig.~\ref{fig:TimingDiagram_d}.
	
	During a UE's session, for $i\geq1$, the $i^{th}$ CRT in a macrocell is denoted by $t_{m_{i}}$ and the $i^{th}$ CRT in a femtocell is denoted by $t_{f_{i}}$. For $i=0$, $t_{m_{0}}$ refers to previous macrocell residence time in which a session starts, and the same for $t_{f_{0}}$. Next, we list the assumptions in our analysis:
	
	\begin{enumerate}
	\item Both macrocell and femtocell CRT, $t_{m_{i}}$ and $t_{f_{i}}$, are i.i.d. and generally distributed as $f_{m}(t)$ and $f_{f}(t)$, with mean $1/\eta_{ m}$ and $1/\eta_{f}$, respectively. Their Laplace transforms are denoted as $f^{*}_{m}(s)$ and $f^{*}_{f}(s)$.
	\item The session length, $t_{s}$, follows exponential distribution with mean $1/\eta_{s}$, and with the probability density function (pdf) $f_{s}(t) = \eta_{s}e^{-\eta_{s}t}$.
	\item The offloading threshold, $t_{o}$, is exponentially distributed with mean $1/\eta_{o}$, and with the pdf $f_{o}(t) = \eta_{o}e^{-\eta_{o}t}$.
	\end{enumerate}

	During a session, we define $t_{s_{i}}$ as the period of time from a UE's $i^{th}$ handover into a femtocell to the time the session ends. Since $t_{s}$ is exponentially distributed, from the memoryless property of exponential process~\cite{ross2014introduction}, $t_{s_{i}}$ has the same distribution as $t_{s}$, i.e., its pdf $f_{s_{i}}(t) = \eta_{s}e^{-\eta_{s}t}$. We summarize the notations used in our analytical model in Table~\ref{tab:ParameterSetting}.
	
	\begin{table}[t]
		\caption{List of Parameters}
		\centering
		\label{tab:ParameterSetting}
		\begin{tabular}{| p{6.5cm}|c|}
			\hline
			                                                              & \textbf{Parameter} \\ \hline\hline
			Session length                                                &       $t_s$        \\ \hline
			Femtocell residence time                                      &      $t_{f}$       \\ \hline
			Macrocell residence time                                      &      $t_{m}$       \\ \hline
			Offloading threshold                                          &      $t_{o}$       \\ \hline
			Offloading time during the session with TO                    &      $T_{t}$       \\ \hline
			Offloading time during the session without TO                 &      $T_{b}$       \\ \hline
			Total number of handovers during the session with TO          &      $N_{t}$       \\ \hline
			Total number of handovers during the session without TO       &      $N_{b}$       \\ \hline
			Number of macrocell-to-femtocell crossings during the session &      $N_{f}$       \\ \hline
			Number of femtocell-to-macrocell crossings during the session &      $N_{m}$       \\ \hline
			Number of macrocell-to-femtocell handovers during the session &      $N_{h}$       \\ \hline
			Residual life of macrocell residence time                     &     $\psi_{m}$     \\ \hline
			Residual life of femtocell residence time                     &     $\psi_{f}$     \\ \hline
			Probability density function of $t_s$                         &      $f_s(t)$      \\ \hline
			Probability density function of $t_{f}$                       &     $f_{f}(t)$     \\ \hline
			Probability density function of $t_{m}$                       &     $f_{m}(t)$     \\ \hline
			Probability density function of $t_{o}$                       &     $f_{o}(t)$     \\ \hline
			Mean session length                                           &    $1/\eta_{s}$    \\ \hline
			Mean femtocell residence time                                 &    $1/\eta_{f}$    \\ \hline
			Mean macrocell residence time                                 &    $1/\eta_{m}$    \\ \hline
			Mean offloading threshold                                     &    $1/\eta_{o}$    \\ \hline
		\end{tabular}
	\end{table}
	
	\subsection{Derivation of Signaling Overhead Reduction Ratio $\Theta(t_{o})$}	
	
	Recall that $\Theta(t_{o}) = (E[N_{b}]- E[N_{t}(t_{o})])/E[N_{b}]$. In the following, we derive $E[N_{b}]$ and $E[N_{t}(t_{o})]$, respectively. There are four possible scenarios for a UE session, as shown in Fig.~\ref{fig:TimingDiagram}. We classify them into two cases: Case~1 for the session starting in a macrocell, as shown in Figs.~\ref{fig:TimingDiagram_a} and~\ref{fig:TimingDiagram_b}, and Case~2 for the session starting in a femtocell, as shown in Figs.~\ref{fig:TimingDiagram_c} and~\ref{fig:TimingDiagram_d}.
%
	The UE stays in femtocell and macrocell alternatively as it moves during a session. From the alternative renewal process~\cite{ross2014introduction}, we can obtain:
	\begin{equation}
		Pr[\text{Case1}] = \frac{E[t_{m_{i}}]}{E[t_{m_{i}}]+E[t_{f_{i}}]} =\frac{\eta_{f}}{\eta_{f}+\eta_{m}} \ ,  \label{case1}
	\end{equation}	
	\begin{equation}
		Pr[\text{Case2}] = \frac{E[t_{f_{i}}]}{E[t_{m_{i}}]+E[t_{f_{i}}]} = \frac{\eta_{m}}{\eta_{f}+\eta_{m}} \ .  \label{case2}
	\end{equation}
	Following the mathematical results of~\cite{fu2013reducing}, we directly have:
	
	\begin{align}
		E[N_{b}] =& \sum_{k=1}^{\infty} k Pr[N_{b}=k] = \frac{2\eta_{m}\eta_{f}}{\eta_{s}(\eta_{m} + \eta_{f})} \ , \label{e_nb} \\
		\alpha =& Pr[t_{o} < t_{f_{i}} | t_{f_{i}} < t_{s_{i}}]
		= \frac{ f^{*}_{f}(\eta_{s}) - f^{*}_{f}(\eta_{s} + \eta_{o}) }{f^{*}_{f}(\eta_{s})} \ , \\
		\beta =& Pr[t_{o} < t_{s_{i}} | t_{s_{i}} < t_{f_{i}} ] = \frac{ (\frac{\eta_{o}}{\eta_{s} + \eta_{o}}) + (\frac{\eta_{s}}{\eta_{s} + \eta_{o}}) f^{*}_{f}(\eta_{s} + \eta_{o}) - f^{*}_{f}(\eta_{s})  }{1- f^{*}_{f}(\eta_{s})} \ ,
	\end{align}
	
	where $\alpha$ and $\beta$ are the conditional probability that a UE handovers into a femtocell during $t_{f_{i}}$, given that the session does not end during this $t_{f_{i}}$ and ends during this $t_{f_{i}}$, respectively. Then, to deride $E[N_{t}(t_{o})]$, we express it as:
		
	\begin{align}
		\nonumber	& E[N_{t}(t_{o})] \\
		=& Pr[\text{Case1}]E[N_{t}(t_{o}) | \text{Case1}] + Pr[\text{Case2}]E[N_{t}(t_{o}) | \text{Case2}] \ .  \label{e_nt_def}
	\end{align}
	%
	%
	
	Let $N_{f}$ and $N_{m}$ be the total number of macrocell-to-femtocell crossings during $t_{s}$ and the total number of femtocell-to-macrocell corssings during $t_{s}$, respectively, where $N_{f} + N_{m} = N_{b}$.
	
	\textbf{Case~1-1}: The number of cell crossings is even, i.e., $N_{1,b} = 2i$, $\forall i \in \mathbb N_{> 0}$, and $N_{f} = N_{m} = \frac{N_{1,b}}{2}$. For Case~1-1, thus, we have:
	\begin{align}
		\nonumber	 & E[N_{t}(t_{o}) | \text{Case1-1} ] \\
		\nonumber  	=& \sum_{i=1}^{\infty}\sum_{k=0}^{i}(i+k) Pr[N_{t}(t_{o})=i+k | N_{1,b}=2i]Pr[N_{1,b}=2i]\\
		\nonumber  	=& \sum_{i=1}^{\infty}\sum_{k=0}^{i} [(i+k) \binom{i}{k} \alpha^{k}(1- \alpha )^{i-k}] Pr[N_{1,b}=2i] \\
		\nonumber  	=& \sum_{i=1}^{\infty} [(1+\alpha)i] (\frac{\eta_{m}}{\eta_{s}})f^{*}_{f}(\eta_{s}) [1-f^{*}_{m}(\eta_{s})]^{2}[f^{*}_{m}(\eta_{s})f^{*}_{f}(\eta_{s})]^{i-1}\\
		=&
		\frac{ \eta_{m}f^{*}_{f}(\eta_{s})[1-f^{*}_{m}(\eta_{s})]^{2} (1+\alpha) }
		{ \eta_{s} (1- f^{*}_{m}(\eta_{s})f^{*}_{f}(\eta_{s}))^{2} } \ .  \label{e_nt_case1_1}
	\end{align}
	
	\textbf{Case~1-2}: The number of cell crossings is odd, i.e., $N_{1,b} = 2i+1$, $\forall i \in \mathbb N$, $N_{f} = \frac{N_{1,b}+1}{2}$ and $N_{m} = \frac{N_{1,b}-1}{2}$. For Case~1-2, thus, we have:
	\begin{align}
		\nonumber	 & E[N_{t}(t_{o}) | \text{Case1-2} ] \\
		\nonumber  	=& \sum_{i=0}^{\infty} \Big [  \beta + \sum_{k=0}^{i} (i+k)Pr[N_{t}(t_{o})=i+k | N_{1,b}=2i+1] \Big] \\
		\nonumber  & \times Pr[N_{1,b}=2i+1]\\
		\nonumber  	=& \sum_{i=0}^{\infty} \Big( \beta + \sum_{k=0}^{i} (i+k) \binom{i}{k} \alpha^{k}(1- \alpha )^{i-k}  \Big) Pr[N_{1,b}=2i+1] \\
		\nonumber  	=& \sum_{i=0}^{\infty} [(1+\alpha)i + \beta] (\frac{\eta_{m}}{\eta_{s}})[1 - f^{*}_{f}(\eta_{s})][1-f^{*}_{m}(\eta_{s})][f^{*}_{m}(\eta_{s})f^{*}_{f}(\eta_{s})]^{i}\\
		=&
		\frac{ \eta_{m} [1 - f^{*}_{f}(\eta_{s})][1-f^{*}_{m}(\eta_{s})] [\beta + (1+\alpha-\beta)f^{*}_{m}(\eta_{s})f^{*}_{f}(\eta_{s})] }
		{ \eta_{s} (1- f^{*}_{m}(\eta_{s})f^{*}_{f}(\eta_{s}))^{2} } \ . \label{e_nt_case1_2}
	\end{align}
	
	\textbf{Case~2-1}: The number of cell crossings is odd, i.e., $N_{2,b} = 2i+1$, $\forall i \in \mathbb N$, $N_{f} = \frac{N_{2,b}-1}{2}$ and $N_{m} = \frac{N_{2,b}+1}{2}$. For Case~2-1, thus, we have:
	\begin{align}
		\nonumber	 & E[N_{t}(t_{o}) | \text{Case2-1} ] \\
		\nonumber  	=& \sum_{i=0}^{\infty} \sum_{k=0}^{i} (i+1+k)Pr[N_{t}(t_{o})=i+k | N_{2,b}=2i+1] \\
		\nonumber     & \times Pr[N_{2,b}=2i+1] \\
		\nonumber  	=& \sum_{i=0}^{\infty} \sum_{k=0}^{i} [(i+1+k) \binom{i}{k} \alpha^{k}(1- \alpha )^{i-k} ] Pr[N_{2,b}=2i+1] \\
		\nonumber  	=& \sum_{i=1}^{\infty} [1 + (1+\alpha)i ] (\frac{\eta_{f}}{\eta_{s}})[1 - f^{*}_{f}(\eta_{s})][1-f^{*}_{m}(\eta_{s})][f^{*}_{m}(\eta_{s})f^{*}_{f}(\eta_{s})]^{i}\\
		=&
		\frac{ \eta_{f} [1 - f^{*}_{f}(\eta_{s})][1-f^{*}_{m}(\eta_{s})] [1 + \alpha f^{*}_{m}(\eta_{s})f^{*}_{f}(\eta_{s})] }
		{ \eta_{s} (1- f^{*}_{m}(\eta_{s})f^{*}_{f}(\eta_{s}))^{2} } \ .  \label{e_nt_case2_1}
	\end{align}
	
	\textbf{Case~2-2}: The number of cell crossings is even, i.e., $N_{2,b} = 2i$, $\forall i \in \mathbb N_{> 0}$, and $N_{f} = N_{m} = \frac{N_{2,b}}{2}$. For Case~2-2, thus, we have:
	\begin{align}
		\nonumber	 & E[N_{t}(t_{o}) | \text{Case2-2}] \\
		\nonumber  	=& \sum_{i=1}^{\infty} \Big( \beta + \sum_{k=0}^{i-1}(i+k) Pr[N_{t}(t_{o})=i+k | N_{2,b}=2i] \Big) \\
		\nonumber    & \times Pr[N_{2,b}=2i] \\
		\nonumber  	=& \sum_{i=1}^{\infty} \Big( \beta + \sum_{k=0}^{i-1} (i+k) \binom{i-1}{k} \alpha^{k}(1- \alpha )^{i-1-k} \Big) Pr[N_{2,b}=2i] \\
		\nonumber  	=& \sum_{i=1}^{\infty} [(i+\alpha)i - \alpha + \beta] (\frac{\eta_{f}}{\eta_{s}})f^{*}_{m}(\eta_{s}) [1-f^{*}_{f}(\eta_{s})]^{2}[f^{*}_{m}(\eta_{s})f^{*}_{f}(\eta_{s})]^{i-1} \\
		=&
		\frac{ \eta_{f}f^{*}_{m}(\eta_{s})[1-f^{*}_{f}(\eta_{s})]^{2} [1+\beta + (\alpha - \beta) f^{*}_{m}(\eta_{s})f^{*}_{f}(\eta_{s}) ] }
		{ \eta_{s} (1- f^{*}_{m}(\eta_{s})f^{*}_{f}(\eta_{s}))^{2} } \ . \label{e_nt_case2_2}
	\end{align}
	
	By applying (\ref{case1}), (\ref{case2}), 
	(\ref{e_nt_case1_1}), (\ref{e_nt_case1_2}), (\ref{e_nt_case2_1}), and (\ref{e_nt_case2_2}) into (\ref{e_nt_def}), we obtain:
	\begin{align}
		\nonumber	 & E[N_{t}(t_{o})] \\
		\nonumber =&
		(1+\alpha)X_{1} + [1+\beta+(1+2\alpha-\beta)f^{*}_{m}(\eta_{s})f^{*}_{f}(\eta_{s})]X_{2} \\
		&+ [1+\beta + (\alpha - \beta) f^{*}_{m}(\eta_{s})f^{*}_{f}(\eta_{s})]X_{3}, \label{e_nt}
	\end{align}
	where
	\begin{align*}
		X_{1} =  \frac{ \eta_{f}\eta_{m}f^{*}_{f}(\eta_{s})[1-f^{*}_{m}(\eta_{s})]^{2} }
		{ \eta_{s}(\eta_{f}+\eta_{m}) (1- f^{*}_{m}(\eta_{s})f^{*}_{f}(\eta_{s}))^{2} } \ , \\
		X_{2} =  \frac{ \eta_{f}\eta_{m} [1-f^{*}_{f}(\eta_{s})] [1-f^{*}_{m}(\eta_{s})] }
		{ \eta_{s}(\eta_{f}+\eta_{m}) (1- f^{*}_{m}(\eta_{s})f^{*}_{f}(\eta_{s}))^{2} } \ , \\
		X_{3} =  \frac{ \eta_{f}\eta_{m}f^{*}_{f}(\eta_{s})[1-f^{*}_{m}(\eta_{s})]^{2} }
		{ \eta_{s}(\eta_{f}+\eta_{m}) (1- f^{*}_{f}(\eta_{s})f^{*}_{m}(\eta_{s}))^{2} } \ .
	\end{align*}
	
	Therefore, our first performance metric $\Theta(t_{o})$ can be obtained by:
	\begin{align}
		\Theta(t_{o}) = \frac{E[N_{b}]- E[N_{t}(t_{o})]}{E[N_{b}]} = \frac{\eta_{s}+\eta_{o}f^{*}_{f}(\eta_{s}+\eta_{o})}{2(\eta_{s}+\eta_{o})}
		\ . \label{theta_to}
	\end{align}	
	
	\subsection{Derivation of Offloading Capability Loss Ratio $\Lambda(t_{o})$}
	
	Recall that $\Lambda(t_{o}) = E[T_{t}(t_{o})] / E[T_{b}]$. We first derive $E[T_{b}]$ for the baseline scheme without TO and then $E[T_{t}(t_{o})]$ for the TO algorithm.
	
	\subsubsection{Derivation of $E[T_{b}]$}
	
	For the baseline scheme, let $\tau$ be the average residual life of $t_{f_{0}}$ during $t_{s}$, i.e., the average offloading time in $t_{f_{0}}$. We then have:
	\begin{align}
		\nonumber  \tau =& E[\psi_{f} | t_{s} > \psi_{f} ]\\
		\nonumber       =& \frac{\int_{t_{\psi_{f}} = 0}^{\infty} \int_{ t_{s} = t_{\psi_{f}}}^{\infty}   t_{\psi_{f}} f_{\psi_{f}}(t_{\psi_{f}})  \eta_{s}e^{-\eta_{s} t_{s} }  dt_{s} dt_{\psi_{f}}}{ Pr[t_{s} > \psi_{f}] } \\
		=& \frac{ E[ t_{\psi_{f}} e^{-\eta_{s} t_{\psi_{f}} } ] }{ f^{*}_{\psi_{f}}(\eta_{s}) }\ .  \label{tau}
	\end{align}

	Let $\sigma$ be the average age for $t_{f_{i}}$ during $t_{s}$, where the session ends in this $t_{f_{i}}$, i.e., the average offloading time in the last $t_{f_{i}}$. According to the residual life theorem~\cite{ross2014introduction}, we have:
	\begin{align}
		\sigma = \tau \ .  \label{sigma}
	\end{align}
	
	Let $\xi$ be the average offloading time in $t_{f_{i}}$, where the session does not start or end in this $t_{f_{i}}$. We then have:
	\begin{align}
		\nonumber	\xi =& E[t_{f} | t_{s} > t_{f}]  \\
		\nonumber	    =& \frac{\int_{t_{f} = 0}^{\infty} \int_{ t_{s} = t_{f}}^{\infty}  t_{f} f_{f}(t_{f})  \eta_{s}e^{-\eta_{s} t_{s} }  dt_{s} dt_{f}}{ Pr[t_{s} > t_{f}] } \\
		=& \frac{ E[ t_{f} e^{-\eta_{s} t_{f} } ] }{ f^{*}_{f}(\eta_{s}) }\ .  \label{xi}
	\end{align}	
	
	Again, $E[T_{b}]$ can be derived from the four scenarios and be expressed as:
	\begin{align}
		E[T_{b}] = Pr[\text{Case1}]E[T_{b} | \text{Case1}] + Pr[\text{Case2}]E[T_{b} | \text{Case2}]  \ .  \label{e_tb_def}
	\end{align}	
	
	\textbf{Case~1-1}: Given $N_{1,b}=2i$, $\forall i \in \mathbb N_{> 0}$, $N_{f} = N_{m} = i$, the total offloading time during $t_{s}$ is $ \sum_{j=1}^{i}t_{f_{j}}$. Thus, we have:
	\begin{align}
		\nonumber	 & E[T_{b} | \text{Case1-1}] \\
		\nonumber  	=& \sum_{i=1}^{\infty} E[(\sum_{j=1}^{i}t_{f_{j}} )| N_{1,b}=2i ]Pr[N_{1,b}=2i]\\
		\nonumber   =& \sum_{i=1}^{\infty} i\xi  (\frac{\eta_{m}}{\eta_{s}})f^{*}_{f}(\eta_{s}) [1-f^{*}_{m}(\eta_{s})]^{2}[f^{*}_{m}(\eta_{s})f^{*}_{f}(\eta_{s})]^{i-1}\\
		=&
		\frac{ \eta_{m}f^{*}_{f}(\eta_{s})[1-f^{*}_{m}(\eta_{s})]^{2} \xi}
		{ \eta_{s} (1- f^{*}_{m}(\eta_{s})f^{*}_{f}(\eta_{s}))^{2} } \ .  \label{e_tb_case1_1}
	\end{align}

	\textbf{Case~1-2}: Given $N_{1,b}=2i+1$, $\forall i \in \mathbb N$, $N_{f} = i+1$ and $N_{m} = i$, the total offloading time during $t_{s}$ is $ ( \sum_{j=1}^{i}t_{f_{j}} )+ t_{s_{i+1}}$. Thus, we have:
	\begin{align}
		\nonumber	 & E[T_{b} | \text{Case1-2}] \\
		\nonumber  	=& \sum_{i=0}^{\infty} E[( \sum_{j=1}^{i}t_{f_{j}} )+ t_{s_{i+1}}| N_{1,b}=2i+1 ]Pr[N_{1,b}=2i+1]\\
		\nonumber   =& \sum_{i=1}^{\infty} (i\xi + \sigma ) (\frac{\eta_{m}}{\eta_{s}})[1 - f^{*}_{f}(\eta_{s})][1-f^{*}_{m}(\eta_{s})][f^{*}_{m}(\eta_{s})f^{*}_{f}(\eta_{s})]^{i}\\
		=&
		\frac{ \eta_{m} [1 - f^{*}_{f}(\eta_{s})][1-f^{*}_{m}(\eta_{s})] [ \sigma + ( \xi - \sigma )f^{*}_{m}(\eta_{s})f^{*}_{f}(\eta_{s})] }
		{ \eta_{s} (1- f^{*}_{m}(\eta_{s})f^{*}_{f}(\eta_{s}))^{2} } \ .  \label{e_tb_case1_2}
	\end{align}
	
	\textbf{Case~2-1}: Given $N_{2,b}=2i+1$, $\forall i \in \mathbb N$, $N_{f} = i$ and $N_{m} = i+1$, the total offloading time during $t_{s}$ is $ \psi_{f} + \sum_{j=1}^{i}t_{f_{j}} $. Thus, we have:
	\begin{align}
		\nonumber	 & E[T_{b} | \text{Case2-1}]Pr[\text{Case2-1}] \\
		\nonumber  	=& \sum_{i=0}^{\infty} E[ \psi_{f} + \sum_{j=1}^{i}t_{f_{j}} | N_{2,b}=2i+1 ]Pr[N_{2,b}=2i+1]\\
		\nonumber   =& \sum_{i=1}^{\infty} ( \tau + i\xi ) (\frac{\eta_{f}}{\eta_{s}})[1 - f^{*}_{f}(\eta_{s})][1-f^{*}_{m}(\eta_{s})][f^{*}_{m}(\eta_{s})f^{*}_{f}(\eta_{s})]^{i}\\
		=&
		\frac{ \eta_{f} [1 - f^{*}_{f}(\eta_{s})][1-f^{*}_{m}(\eta_{s})] [ \tau + ( \xi - \tau )f^{*}_{m}(\eta_{s})f^{*}_{f}(\eta_{s})] }
		{ \eta_{s} (1- f^{*}_{m}(\eta_{s})f^{*}_{f}(\eta_{s}))^{2} } \ .  \label{e_tb_case2_1}
	\end{align}
	
	\textbf{Case~2-2}: Given $N_{b}=2i$, $\forall i \in \mathbb N_{> 0}$, $N_{f} = N_{m} = i$, the total offloading time during $t_{s}$ is $ \psi_{f} + (\sum_{j=1}^{i-1}t_{f_{j}}) + t_{s_{i}}$. Thus, we have:
	\begin{align}
		\nonumber	 & E[T_{b} | \text{Case2-2}]Pr[\text{Case2-2}] \\
		\nonumber  	=& \sum_{i=1}^{\infty} E[ \psi_{f} + (\sum_{j=1}^{i-1}t_{f_{j}}) + t_{s_{i}} | N_{2,b}=2i ]Pr[N_{2,b}=2i]\\
		\nonumber   =& \sum_{i=1}^{\infty} [ \tau + (i-1)\xi + \sigma ]  (\frac{\eta_{f}}{\eta_{s}})f^{*}_{m}(\eta_{s})[1-f^{*}_{f}(\eta_{s})]^{2} \\
		\nonumber &\times [f^{*}_{m}(\eta_{s})f^{*}_{f}(\eta_{s})]^{i-1} \\
		=&
		\frac{ \eta_{f}f^{*}_{f}(\eta_{s})[1-f^{*}_{m}(\eta_{s})]^{2} [\tau + \sigma + (\xi - \tau -\sigma)f^{*}_{m}(\eta_{s})f^{*}_{f}(\eta_{s})]}
		{ \eta_{s} \eta_{f} (1- f^{*}_{m}(\eta_{s})f^{*}_{f}(\eta_{s}))^{2} } \ .  \label{e_tb_case2_2}
	\end{align}
	
	By applying (\ref{case1}), (\ref{case2}), (\ref{e_tb_case1_1}), (\ref{e_tb_case1_2}), (\ref{e_tb_case2_1}), and (\ref{e_tb_case2_2}) into (\ref{e_tb_def}), we obtain:
	\begin{align}
		\nonumber	 & E[ T_{b} ] \\
		\nonumber   =&
		\xi X_{1} + [ \tau + \sigma +( 2\xi - \tau - \sigma)f^{*}_{m}(\eta_{s})f^{*}_{f}(\eta_{s})]X_{2} \\
		&+ [\tau + \sigma + (\xi - \tau -\sigma)f^{*}_{m}(\eta_{s})f^{*}_{f}(\eta_{s})]X_{3}, \label{e_tb}
	\end{align}
	where $X_{1}$, $X_{2}$, and $X_{3}$ are the same as those in Eq. (\ref{e_nt}).

	\subsubsection{Derivation of $E[T_{t}(t_{o})]$}		
	For the TO algorithm, let $\phi$ be the average offloading time for $t_{f_{i}}$ during $t_{s}$, where the session does not start or end in this $t_{f_{i}}$. We then have:
	\begin{align}
		\nonumber	 \phi =& E[t_{f_{i}} - t_{o} | ( t_{o} < t_{f_{i}} | t_{f_{i}} < t_{s_{i}} ) ]  \\
		\nonumber         =&  \int_{t_{o}=0}^{\infty} \int_{t_{f_{i}} = t_{o}}^{\infty} \int_{ t_{s_{i}} = t_{f_{i}}}^{\infty} (t_{f_{i}} - t_{o}) \\
		\nonumber &\times  \frac{\eta_{o}e^{-\eta_{o}t_{o}} f_{f}(t_{f_{i}}) \eta_{s}e^{-\eta_{s} t_{s_{i}} }}{Pr[t_{o} < t_{f_{i}} | t_{f_{i}} < t_{s_{i}}] Pr[ t_{f_{i}} < t_{s_{i}} ]}  dt_{s_{i}} dt_{f_{i}} dt_{o} \\
		=&  \frac{ E[t_{f_{i}}e^{- \eta_{s}t_{f_{i}}}] + \frac{1}{\eta_{o}}(f^{*}_{f}(\eta_{s}+\eta_{o}) - f^{*}_{f}(\eta_{s})) }{\alpha f^{*}_{f}(\eta_{s})}  \ .
	\end{align}
	
	Let $\rho$ be the average offloading time for $t_{f_{i}}$ during $t_{s}$, where the session ends in this $t_{f_{i}}$.	We then have:
	\begin{align}
		\nonumber	 &\rho = E[t_{s_{i}} - t_{o} | (t_{o} < t_{s_{i}} | t_{s_{i}} < t_{f_{i}} ) ]  \\
		\nonumber         =&  \int_{t_{o}=0}^{\infty} \int_{t_{s_{i}} = t_{o}}^{\infty} \int_{ t_{f_{i}} = t_{s_{i}}}^{\infty} (t_{s_{i}} - t_{o}) \\
		\nonumber          & \times
		\frac{\eta_{o}e^{-\eta_{o}t_{o}} \eta_{s}e^{-\eta_{s} t_{s_{i}} } f_{f}(t_{f_{i}}) }{Pr[t_{o} < t_{s_{i}} | t_{s_{i}} < t_{f_{i}} ] Pr[t_{s_{i}} < t_{f_{i}}] }  dt_{f_{i}} dt_{s_{i}} dt_{o} \\
		=&  \frac{ \frac{1}{\eta_{s}} + (\frac{1}{\eta_{o}} - \frac{1}{\eta_{s}})f^{*}_{f}(\eta_{s}) -E[t_{f_{i}}e^{- \eta_{s}t_{f_{i}}}] - \frac{\eta_{o}+\eta_{s} f^{*}_{f}(\eta_{s}+\eta_{o})}{\eta_{o}(\eta_{s}+\eta_{o})} }{ \beta (1 - f^{*}_{f}(\eta_{s}))}\ .
	\end{align}
	
	Similar to the derivation of $E[T_{b}]$, we express $E[T_{t}(t_{o})]$ as:
	\begin{align}
		\nonumber 	& E[T_{t}(t_{o})] \\
		=& Pr[\text{Case1}]E[T_{t}(t_{o}) | \text{Case1}] + Pr[\text{Case2}]E[T_{t}(t_{o}) | \text{Case2}] \ .  \label{e_to_def}
	\end{align}
	
	Let $N_{h}$ be the total number of macrocell-to-femtocell handovers during $t_{s}$. $E[T_{t}(t_{o})]$ is derived from the four scenarios as follows: 	
	
	\textbf{Case~1-1}: Given $N_{1,b}=2i$, $N_{f} = i$, $\forall i \in \mathbb N_{> 0}$, $N_{h}$ is in the range $0 \leq N_{h} \leq i$. The average offloading time during $t_{s}$ is ($\sum_{j=0}^{i}j\phi Pr[N_{h}=j]$). Thus, we have:
	\begin{align}
		\nonumber	 & E[T_{t}(t_{o}) | \text{Case1-1} ] \\
		\nonumber  	=& \sum_{i=1}^{\infty} \sum_{k=0}^{i} (k\phi) Pr[N_{h}=k | N_{1,b}=2i]Pr[N_{1,b}=2i]\\
		\nonumber  	=& \sum_{i=1}^{\infty} \sum_{k=0}^{i} [(k\phi) \binom{i}{k} \alpha^{k}(1- \alpha )^{i-k}] Pr[N_{1,b}=2i] \\
		\nonumber  	=& \sum_{i=1}^{\infty} [\alpha \phi i] (\frac{\eta_{m}}{\eta_{s}})f^{*}_{f}(\eta_{s}) [1-f^{*}_{m}(\eta_{s})]^{2}[f^{*}_{m}(\eta_{s})f^{*}_{f}(\eta_{s})]^{i-1}\\
		=&
		\frac{ \eta_{m}f^{*}_{f}(\eta_{s})[1-f^{*}_{m}(\eta_{s})]^{2} (\alpha \phi) }
		{ \eta_{s} (1- f^{*}_{m}(\eta_{s})f^{*}_{f}(\eta_{s}))^{2} } \ .  \label{e_to_case1_1}
	\end{align}
	
	\textbf{Case~1-2}: Given $N_{1,b}=2i+1$, $N_{f} = i+1$, $\forall i \in \mathbb N$, $N_{h}$ is in the range $1 \leq N_{h} \leq i+1$. The average offloading time during $t_{s}$ is ($\beta \rho + \sum_{j=0}^{i}j\phi Pr[N_{h}=j+1]$). Thus, we have:
	\begin{align}
		\nonumber	 & E[T_{t}(t_{o}) | \text{Case1-2} ] \\
		\nonumber  	=& \sum_{i=0}^{\infty} \Big( \beta\rho + \sum_{k=0}^{i} (k\phi) Pr[N_{h}=k+1 | N_{1,b}=2i+1]  \Big)  \\
		\nonumber    &\times Pr[N_{1,b}=2i+1] \\
		\nonumber  	=& \sum_{i=0}^{\infty} \Big( \beta\rho + \sum_{k=0}^{i} (k\phi) \binom{i}{k} \alpha^{k}(1- \alpha )^{i-k} \Big) \\
		\nonumber    &\times Pr[N_{1,b}=2i+1] \\
		\nonumber  	=& \sum_{i=0}^{\infty} [\alpha \phi i + \beta\rho] (\frac{\eta_{m}}{\eta_{s}})[1 - f^{*}_{f}(\eta_{s})][1-f^{*}_{m}(\eta_{s})] \\
		\nonumber    &\times [f^{*}_{m}(\eta_{s})f^{*}_{f}(\eta_{s})]^{i}\\
		=&
		\frac{ \eta_{m} [1 - f^{*}_{f}(\eta_{s})][1-f^{*}_{m}(\eta_{s})] [\beta\rho + (\alpha\phi-\beta\rho)f^{*}_{m}(\eta_{s})f^{*}_{f}(\eta_{s})] }
		{ \eta_{s} (1- f^{*}_{m}(\eta_{s})f^{*}_{f}(\eta_{s}))^{2} } \ .  \label{e_to_case1_2}
	\end{align}

	\textbf{Case~2-1}: Given $N_{2,b}=2i+1$, $N_{f} = i$, $\forall i \in \mathbb N$, $N_{h}$ is in the range $0 \leq N_{h} \leq i$. The average offloading time during $t_{s}$ is ($\tau + \sum_{j=0}^{i}j\phi Pr[N_{h}=j]$). Thus, we have:
	\begin{align}
		\nonumber	 & E[T_{t}(t_{o}) | \text{Case2-1} ] \\
		\nonumber  	=& \sum_{i=0}^{\infty} \Big( \tau + \sum_{k=0}^{i} (k\phi) Pr[N_{h}=k | N_{2,b}=2i+1] \Big) \\
		\nonumber     &\times Pr[N_{2,b}=2i+1]\\
		\nonumber  	=& \sum_{i=0}^{\infty} \Big( \tau + \sum_{k=0}^{i} (k\phi) \binom{i}{k} \alpha^{k}(1- \alpha )^{i-k} \Big) \\
		\nonumber     &\times Pr[N_{2,b}=2i+1]\\
		\nonumber  	=& \sum_{i=0}^{\infty} [\tau + \alpha \phi i ] (\frac{\eta_{f}}{\eta_{s}})[1 - f^{*}_{f}(\eta_{s})][1-f^{*}_{m}(\eta_{s})][f^{*}_{m}(\eta_{s})f^{*}_{f}(\eta_{s})]^{i}\\
		=&
		\frac{ \eta_{f} [1 - f^{*}_{f}(\eta_{s})][1-f^{*}_{m}(\eta_{s})] [\tau + (\alpha\phi-\tau)f^{*}_{m}(\eta_{s})f^{*}_{f}(\eta_{s})] }
		{ \eta_{s} (1- f^{*}_{m}(\eta_{s})f^{*}_{f}(\eta_{s}))^{2} } \ .  \label{e_to_case2_1}
	\end{align}
	
	\textbf{Case~2-2}: Given $N_{2,b}=2i$, $N_{f} = i$, $\forall i \in \mathbb N_{> 0}$, $N_{h}$ is in the range $1 \leq N_{h} \leq i$. The average offloading time during $t_{s}$ is ($\tau + \beta \rho + \sum_{j=0}^{i-1}j\phi Pr[N_{h}=j+1]$). Thus, we have:
	
	\begin{align}
		\nonumber	 & E[T_{t}(t_{o}) | \text{Case2-2} ] \\
		\nonumber  	=& \sum_{i=1}^{\infty} \Big( \tau + \beta\rho + \sum_{k=0}^{i-1} (k\phi) Pr[N_{h}=k+1 | N_{2,b}=2i] \Big)  Pr[N_{2,b}=2i]\\
		\nonumber  	=& \sum_{i=1}^{\infty} \Big( \tau + \beta\rho + \sum_{k=0}^{i-1} (k\phi) \binom{i-1}{k} \alpha^{k}(1- \alpha )^{i-1-k} \Big) Pr[N_{2,b}=2i]\\
		\nonumber  	=& \sum_{i=1}^{\infty} [\tau + \beta\rho + \alpha \phi (i-1)] (\frac{\eta_{f}}{\eta_{s}})f^{*}_{m}(\eta_{s}) [1-f^{*}_{f}(\eta_{s})]^{2}  [f^{*}_{m}(\eta_{s})f^{*}_{f}(\eta_{s})]^{i-1}\\
		\nonumber   =&
		\frac{ \eta_{m}f^{*}_{m}(\eta_{s})[1-f^{*}_{f}(\eta_{s})]^{2} [\tau + \beta\rho ] }
		{ \eta_{s} (1- f^{*}_{m}(\eta_{s})f^{*}_{f}(\eta_{s}))^{2} } \\
		&+ \frac{ \eta_{m}f^{*}_{m}(\eta_{s})[1-f^{*}_{f}(\eta_{s})]^{2} [(\alpha \phi - \tau - \beta\rho)f^{*}_{m}(\eta_{s})f^{*}_{f}(\eta_{s})] }
		{ \eta_{s} (1- f^{*}_{m}(\eta_{s})f^{*}_{f}(\eta_{s}))^{2} } \ .  \label{e_to_case2_2}
	\end{align}
	
	By applying (\ref{case1}), (\ref{case2}), (\ref{e_to_case1_1}), (\ref{e_to_case1_2}), (\ref{e_to_case2_1}), and (\ref{e_to_case2_2}) into (\ref{e_to_def}), we have:
	\begin{align}
		\nonumber	 & E[ T_{t}(t_{o}) ] \\
		\nonumber   =&
		\alpha \phi X_{1} + [ \tau + \beta\rho +(2\alpha \phi - \tau - \beta\rho )f^{*}_{m}(\eta_{s})f^{*}_{f}(\eta_{s})]X_{2} \\
		&+  [\tau + \beta\rho + (\alpha \phi - \tau - \beta\rho )f^{*}_{m}(\eta_{s})f^{*}_{f}(\eta_{s})]X_{3}, \label{e_to}
	\end{align}
	where $X_{1}$, $X_{2}$, and $X_{3}$ are the same as those in Eq. (\ref{e_nt}).
	
	Therefore, our second performance metric $\Lambda(t_{o})$ can be obtained by:
	\begin{align}
		\Lambda(t_{o}) = \frac{E[T_{t}(t_{o})]}{E[T_{b}]} = \frac{\eta_{f}\eta_{o}-\eta_{f}(\eta_{s}+\eta_{o})f^{*}_{f}(\eta_{s})+\eta_{f}\eta_{s}f^{*}_{f}(\eta_{s}+\eta_{o})+\eta^{2}_{s}(\eta_{s}+\eta_{o})Y_{2}}{\eta_{s}(\eta_{s}+\eta_{o})(\eta_{f}Y_{1}+2\eta_{s}Y_{2})}
		\ .   \label{lambda_to}
	\end{align}	
	where
	$Y_{1} = E[ t_{f} e^{-\eta_{s} t_{f} } ]$ and $Y_{2} = E[ t_{\psi_{f}} e^{-\eta_{s} t_{\psi_{f}} } ]$ .

	\section{Simulation and Numerical Results}
	\label{simulationandnumericalresults}
	
	In this section, we provide numerical results for the analysis presented in Section~\ref{analyticalmodel}. The analysis is validated through extensive simulations by using ns-2~\cite{ns2}, version 2.35. In addition, the errors between the analytical and simulated results are verified to fall within one percent. Due to page limitation, we only show one table of our results, Table~\ref{tab:Validation}.

	The two performance metrics, i.e., (\ref{theta_to}) and (\ref{lambda_to}), can be organized into a meaningful form only if $f^{*}_{f}(s)$ and $f^{*}_{m}(s)$ have closed-form solutions. Thus, we choose to apply Gamma distribution for both femtocell residence time and macrocell residence time because the Gamma distribution has two characteristics: (1) it is a more general kind of distribution and can approximate many other distributions, and (2) a closed-form Laplace transform solution for it exists. Also, the Gamma distribution has been commonly adopted in many previous papers (e.g.~\cite{ma2004dynamic,liou2013investigation}) to study various UE mobility behaviors.

\begin{figure*}[t]
		\graphicspath{ {./Figures/} }
		\centering
		\includegraphics[width=15cm]{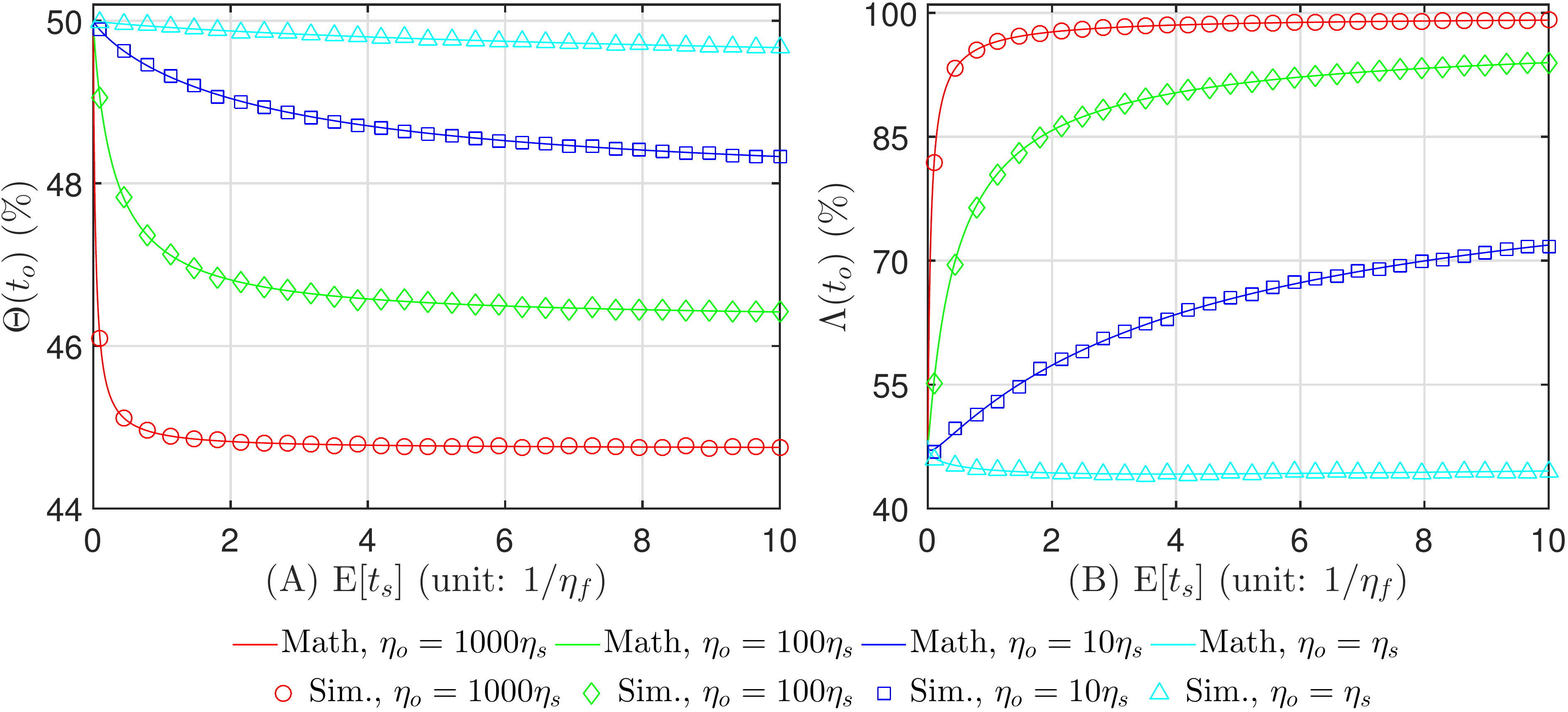}
		\caption{The effects of mean session length $\eta_{s}$ on $\Theta(t_{o})$ and $\Lambda(t_{o})$ ($1/\eta_{m} = 60s$, $1/\eta_{f} = 15s$).}
		\label{fig:Effects_eta_s}
		\vspace{-10mm}
	\end{figure*}

	Then, we observe that two factors affects the performance of our proposed TO: the effects of UE mobility and the session length. Their impact are investigated and the results are illustrated in Figs.~\ref{fig:Effects_eta_f}\textendash\ref{fig:Effects_eta_s}.
	
	\begin{table*}[t]
		\caption{Validation of analytical and simulated results.}
		\centering
		\label{tab:Validation}
		\begin{tabular}{ |c|c|c|c|c|c| }
			\hline
			\multicolumn{6}{ |c| }{ $\eta_m=10\eta_s$, $v_{m}=1/\eta_{m}$, $\eta_f=10\eta_s$, $v_{f}=1000/\eta_f$ }  \\ \hline
			\multicolumn{2}{ |c| }{ $1/\eta_{o}$ (unit:$1/\eta_s$)} &  $0.1$  &  $0.2$  &  $0.3$  &  $0.4$  \\ \hline 	\hline
			\multirow{3}{*}{$N_{t}(t_{o})$}   &          Analytical          & 5.74007 & 5.55835 & 5.45672 & 5.38896 \\ \cline{2-6}
			&          Simulated           & 5.75666  & 5.53744  & 5.45196  & 5.38116  \\ \cline{2-6}
			&          Error (\%)          &   0.28899   &   0.37618   &   0.08718   &   0.14468   \\ \hline
			\multirow{3}{*}{$T_{t}(t_{o})$}   &          Analytical          & 192.43505  & 169.83760  & 154.62965  & 143.48846  \\ \cline{2-6}
			&          Simulated           & 191.97960  & 168.85753  & 155.70437  & 142.10245  \\ \cline{2-6}
			&          Error (\%)          &   0.23688   &   0.57707   &   0.69503   &   0.96593   \\ \hline
			\multirow{3}{*}{$\Theta(t_{o})$}  &          Analytical          & 42.59933  & 44.41655  & 45.43281  & 46.11039  \\ \cline{2-6}
			&          Simulated           & 42.59930  & 44.42310  & 45.40490  & 46.12440  \\ \cline{2-6}
			&          Error (\%)          &   0.00008   &   0.01476   &   0.06143   &   0.03038   \\ \hline
			\multirow{3}{*}{$\Lambda(t_{o})$} &          Analytical          & 81.25140  & 71.71013  & 65.28891  & 60.58480  \\ \cline{2-6}
			&          Simulated           & 81.14890  & 71.63550  & 65.33060  & 60.46900  \\ \cline{2-6}
			&            Error (\%)        &  0.12615  &  0.10408  &  0.06385  &  0.19113  \\ \hline
		\end{tabular}
	\end{table*}

	\subsection{Effects of UE Mobility}
	
	As shown in Eqs.~(\ref{theta_to}) and (\ref{lambda_to}), the UE's behavior inside the macrocell ($\eta_{m}$ and $v_{m}$) have no impacts on the two performance metrics, the effects of UE mobility can be simply reflected by the UE's behavior inside femtocells, i.e., $\eta_{f}$ and $v_{f}$.
	
\begin{figure*}[t]
		\vspace{-3mm}
		\graphicspath{ {./Figures/} }
		\centering
		\includegraphics[width=15cm]{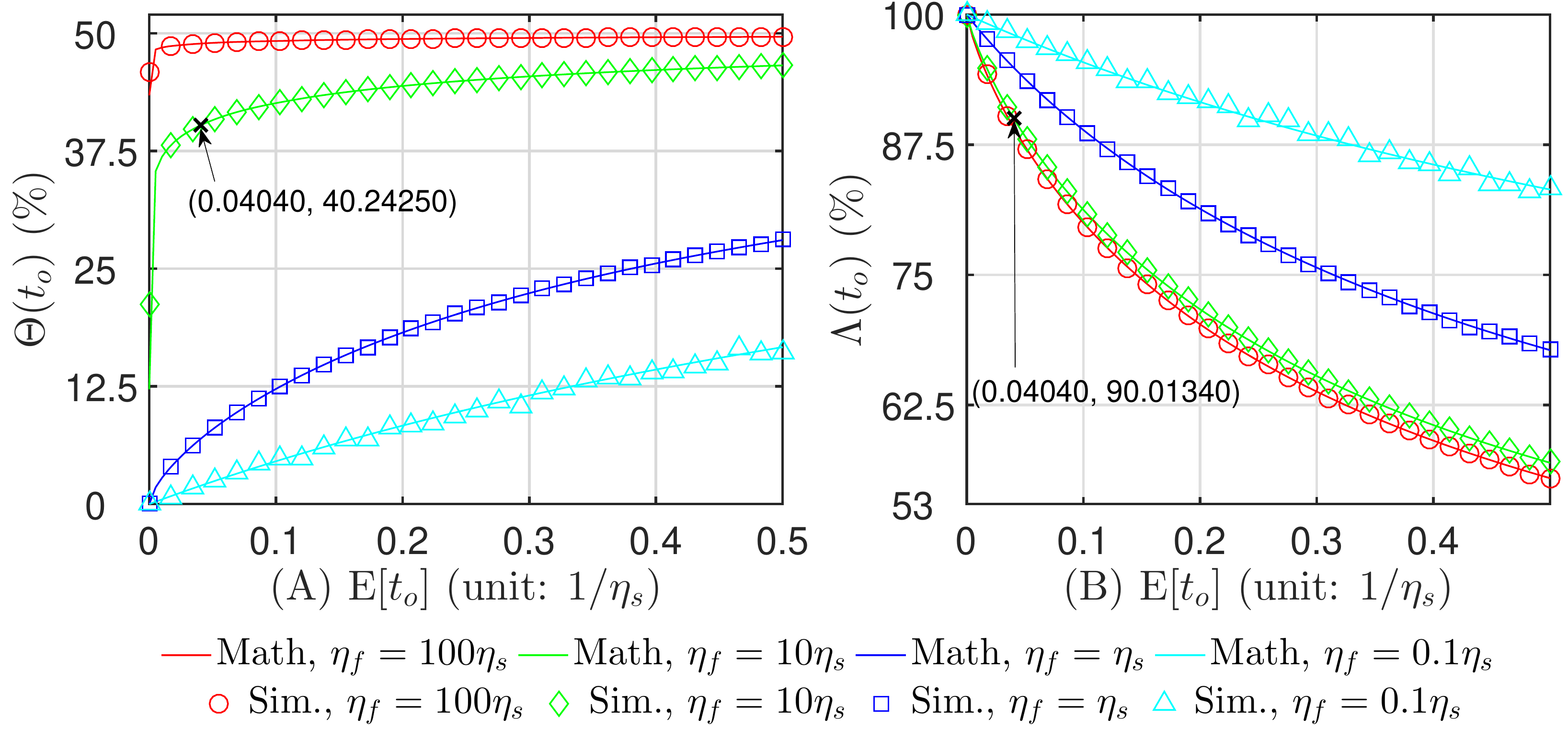}
		\caption{The effects of mean femtocell residence time $\eta_{f}$ on $\Theta(t_{o})$ and $\Lambda(t_{o})$ ($\eta_{m} = 10\eta_{s}$).}
		\label{fig:Effects_eta_f}
		\vspace{-4mm}
	\end{figure*}

	First, the effects of mean femtocell residence time (FRT), $1/\eta_{f}$, are investigated. In Fig.~\ref{fig:Effects_eta_f}, the mean macrocell residence time (MRT), $1/\eta_{m}$, is fixed as one tenth of the session length ($\eta_{m} = 10\eta_{s}$), and the mean FRT is set from one hundredth of the session length ($\eta_{f} = 100\eta_{s}$) to ten times of that ($\eta_{f} = 0.1\eta_{s}$). Generally, for the cases that the mean FRT take smaller portion of the session length, the proposed TO has better performance in terms of $\Theta(t_{o})$. For example, in the green-diamond line, the proposed TO can easily reduce over 40\% signaling overhead but lose less than 10\% femtocell offloading capability. In addition, $\Theta(t_{o})$ increases faster than $\Lambda(t_{o})$ drops as $E[t_{o}]$ rises. Thus, the proposed TO is proven to be capable of significantly reducing signaling overhead at a minor cost of femtocell offloading capability.
	
\begin{figure*}[t]
		\graphicspath{ {./Figures/} }
		\centering
		\includegraphics[width=15cm]{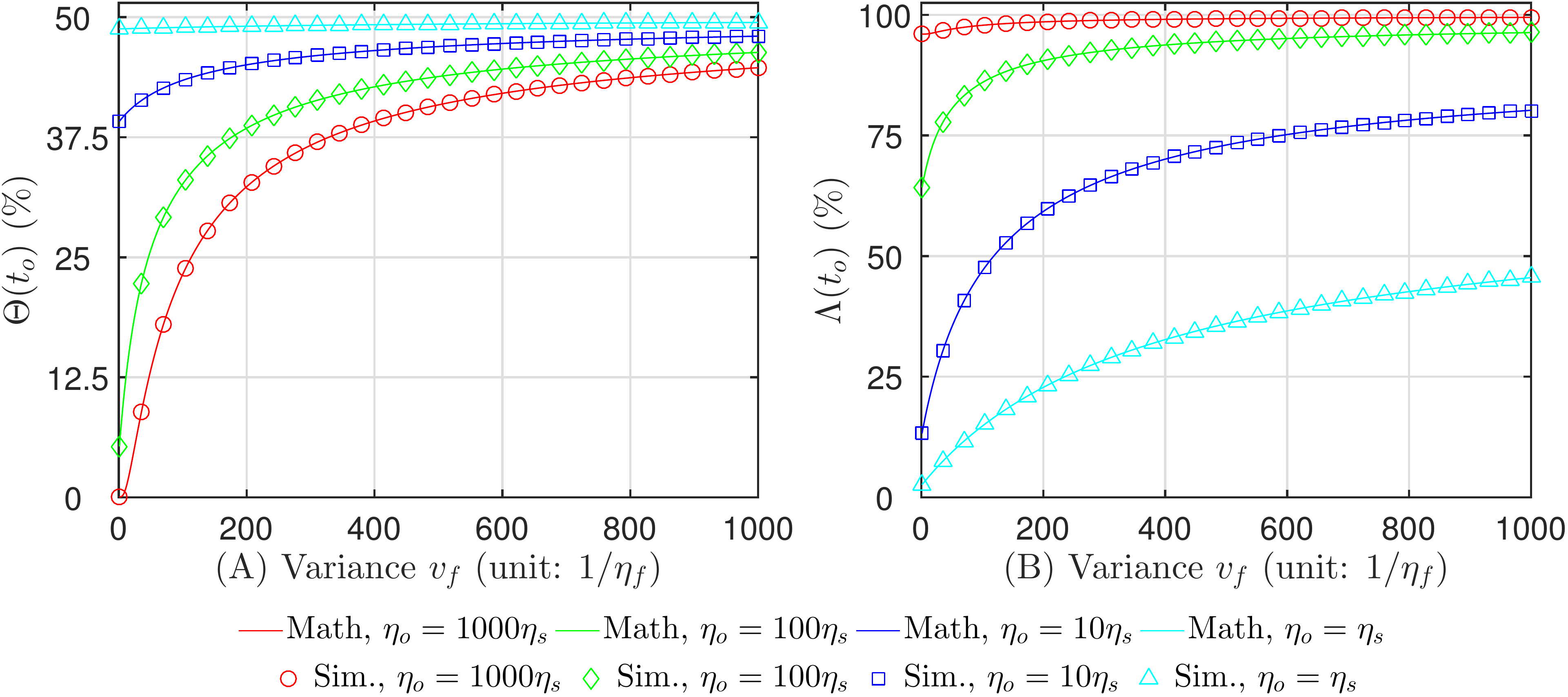}
		\caption{The effects of variance of femtocell residence time $v_{f}$ on $\Theta(t_{o})$ and $\Lambda(t_{o})$ ($\eta_{m} = 10\eta_{s}$, $\eta_{f} = 40\eta_{s}$).}
		\label{fig:Effects_v_f}
		\vspace{-10mm}
	\end{figure*}

	Second, the effects of the variance of FRT, $v_{f}$, are investigated. A larger variance indicates that the UE mobility behavior is more dynamic and less predictable, which fits closer to real life situation. In Fig.~\ref{fig:Effects_v_f}, the mean MRT and the mean FRT are fixed as one tenth of the session length
	($\eta_{m} = 10\eta_{s}$) and one fortieth of the session length ($\eta_{f} = 40\eta_{s}$), respectively. The variance $v_{f}$ is set from one hundredth of mean FRT ($v_{f} = 1/100\eta_{f}$) to a thousand times  of mean FRT ($v_{f} = 1000/\eta_{f}$). As shown in Fig.~\ref{fig:Effects_v_f}, both $\Theta(t_{o})$ and $\Lambda(t_{o})$ have better performance when the variance $v_{f}$ gets larger for all different mean values of the offloading threshold. Therefore, the proposed TO is considered effective against various UE mobility behaviors in real life.
	
	\subsection{Effects of Session Length}
	In Fig.~\ref{fig:Effects_eta_s}, the mean MRT and the mean FRT are fixed as 60 and 15 seconds, respectively. The mean session length is set from 0 to 150 seconds (i.e., $10/\eta_{f}$). We observe that the mean session length has prominent effects on the two performance metrics only when it is smaller than the mean FRT, i.e., $\eta_{s} > \eta_{f}$, while it tends to be less perceivable when $\eta_{s} < \eta_{f}$. In the case of $\eta_{s} > \eta_{f}$, $\Theta(t_{o})$ decreases when the session length increases. It is because the offloading threshold takes smaller proportion of the session length when $t_{s}$ gets larger. On the other hand, for $\eta_{s} > \eta_{f}$, $\Lambda(t_{o})$ increases when the session length increases because the offloading time lengthens along with the session length. Thus, depending on whether the mean session length is smaller than the mean FRT, the significance of the session length can be determined. Overall, for $\eta_{s} > \eta_{f}$, both UE mobility behaviors and the session length have impacts on the two performance metrics, while the effects of UE mobility dominate for the case of $\eta_{s} < \eta_{f}$.

	\section{Optimal TO Algorithm}
	\label{optimaltomechanism}
	
	\subsection{Determination of Optimal Offloading Threshold}
	
	It is of high priority for network operators to select a proper offloading threshold $t_{o}$, such that the TO algorithm is efficient. A large offloading threshold $t_{o}$ will reduce massive network signaling overhead but sacrifice femtocell offloading capability. On the other hand, a small offloading threshold $t_{o}$ reserves more femtocell offloading capability but costs higher network signaling overhead. Recall that $t_{o}$ is an exponential random variable, and our goal is to find the optimal mean offloading threshold, aka the optimal distribution, such that $\Theta(t_{o})$ and $\Lambda(t_{o})$ together achieve maximum. From Eqs. (\ref{theta_to}) and (\ref{lambda_to}), both $\Theta(t_{o})$ and $\Lambda(t_{o})$ are functions of $\eta_{o}$, thus we formulate the objective function as:
		
	\begin{equation}
	\begin{aligned}
	& \underset{\eta_{o}}{\text{maximize}}
	& & f(\eta_{o}) = \Theta(\eta_{o}) + \Lambda(\eta_{o}) \\
	& \text{subject to}
	& & 0 < \eta_{o} \leq \delta \ , \\
	\end{aligned}
	\end{equation}
	
	where $\delta$ is the upperbound of $\eta_{o}$, which can be determined by operators according to their historical statistic data, and define the optimal offloading threshold as:
	
	\begin{equation}
		E[t_{o}]^{*} = \frac{1}{\eta^{*}_{o}} \ . \label{optimal_eta_o}
	\end{equation}
	
	Then, according to Calculus, such $\eta^{*}_{o}$ can be found by solving the differential equation $f^{'}(\eta_{o})~=~0$ and checking the boundary values, i.e., $f(0)$ and $f(\delta)$. Since the closed-form solution to the differential equation does not have a meaningful and simplified expression, we instead acquire the answer by using MATLAB. The mean computing time for the $\eta^{*}_{o}$ obtained by using a normal PC (Intel Core i5-3470 with 8G RAM) is 121.56~ms with the standard deviation of 1.97~ms.

	\begin{figure}[t]
		\graphicspath{ {./Figures/} }
		\centering
		\includegraphics[width=8.7cm]{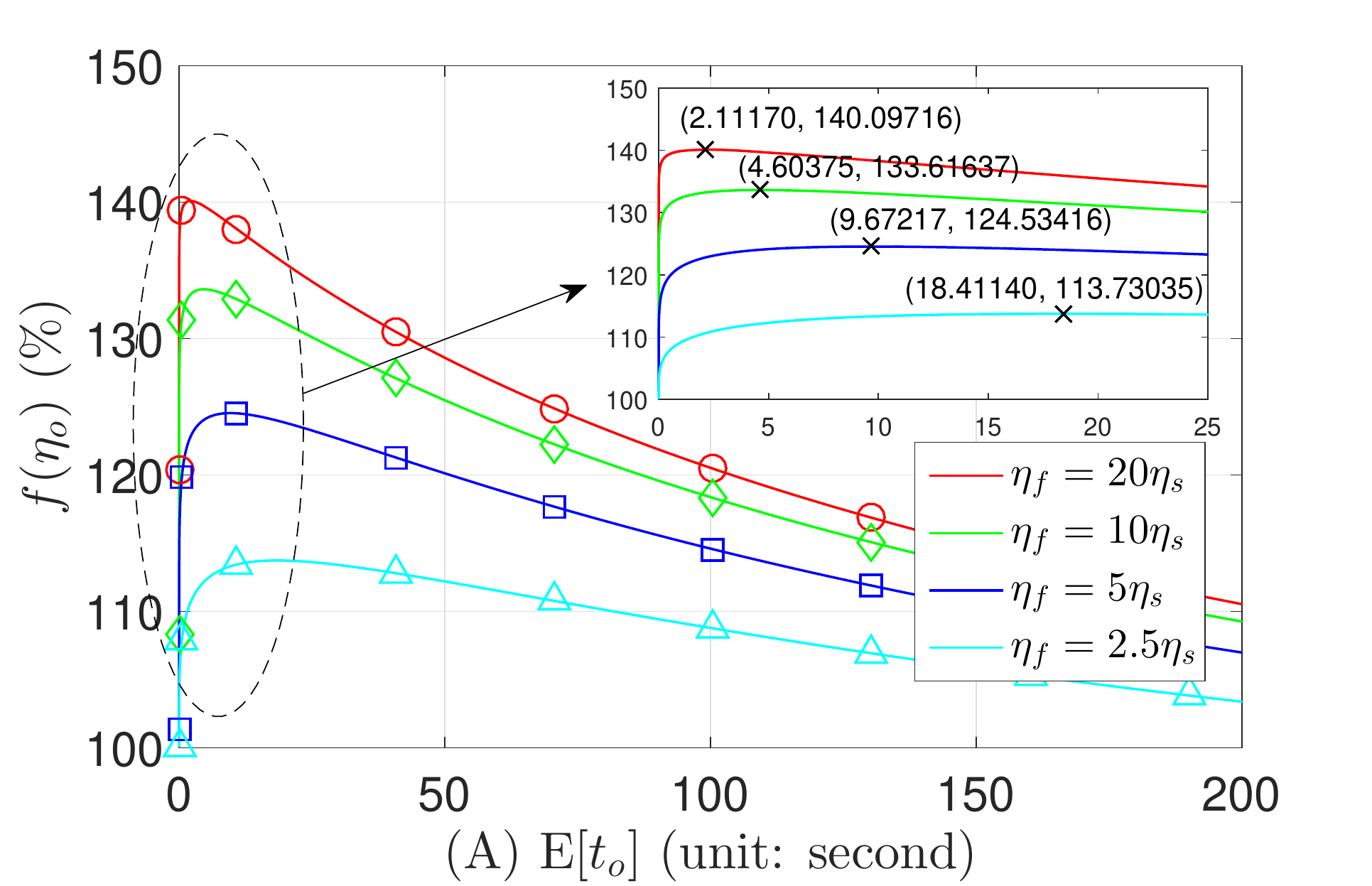}
		\caption{Determination of the optimal offloading threshold.}
		\label{fig:optimal_to}
		\vspace{-7mm}
	\end{figure}	 	
	
	Fig.~\ref{fig:optimal_to} illustrates a graphical plot of $f(\eta_{o})$ with 4 different settings. In this example, the session length $t_{s}$ and the upperbound $\delta$ are set as 600 second and 200 second, respectively. We solve the equation $f^{'}(\eta_{o}) = 0$ and compare it with the boundary values for each line, and the maximum values are denoted as crosses in the figure. For the green line ($\eta_{f}=10\eta_{s}$), the optimal offloading threshold $E[t_{o}]^{*}$ is computed as 4.60375, with the corresponding $\eta^{*}_{o}$ as 0.21721. Thus, operators can achieve the best performance for the TO algorithm by setting the optimal $\eta^{*}_{o}$.
	
	Next, we propose an algorithm (shown as Algorithm~\ref{algo}) to systematically find $\eta^{*}_{o}$. Algorithm~\ref{algo} takes one input value: $\delta$, which is the upperbound for $\eta_{o}$. Again, operators can refer to their historical statistics on femtocell CRT and determine $\delta$ accordingly. First, Algorithm~\ref{algo} checks if $f^{'}(\eta_{o})=0$ has a real solution, say, $\eta^{'}_{o}$. If no, then return the larger boundary value. Otherwise, it compares $\eta^{'}_{o}$ with the two boundary values, and return whichever the largest as the optiaml offloading threshold. Note that when $\eta^{*}_{o}$ is returned as 0, $E[t_{o}]^{*}$ would goes to infinity. In practice, this can be solved by just setting $\eta^{*}_{o}$ as an arbitrary small number.

	\begin{algorithm}[t]
		\caption{Selecting an optimal offloading threshold.}
		\label{algo}
		\textbf{Input:} $\delta$ \\
		\textbf{Output:} $\eta^{*}_{o}$
		\begin{algorithmic}[1]
			\ENSURE $\eta^{*}_{o} \in (0, \delta]$
			\IF{$f^{'}(\eta_{o})=0$ has a real solution $\eta^{'}_{o}$}
				\IF{$f(\delta) > f(0)$}
					\IF{$f(\eta^{'}_{o}) > f(\delta)$}
						\RETURN $\eta^{'}_{o}$
					\ELSE
						\RETURN $f(\delta)$
					\ENDIF				
				\ELSE
					\IF{$f(\eta^{'}_{o}) > f(0)$}
						\RETURN $\eta^{'}_{o}$
					\ELSE
						\RETURN $f(0)$
					\ENDIF				
				\ENDIF
			\ELSIF{$f(\delta) > f(0)$}
			\RETURN $\delta$
			\ELSE
			\RETURN $0$
			\ENDIF
		\end{algorithmic}
	\end{algorithm}
	
	\subsection{Implementation Issues}
	
	To use the TO algorithm, we need to gather statistics of $E[t_f]$, $E[t_m]$, and $E[t_s]$ in advance. First, HeNBs can continuously collect each UE's duration in the femtocells by observing the time stamps when it performs handover in and handover out. The HeNBs then upload their own $E[t_f]$ to the eNB, which in turn will compute a combined $E[t_f]$ from all HeNBs. Note that the frequency for uploading $E[t_f]$ involves the tradeoff between the precision of the TO algorithm and extra signaling overhead. Second, eNBs can obtain $E[t_m]$ when HeNBs obtain $E[t_f]$. Last, $E[t_s]$ can be acquired by recording the lifetime of traffic flows at the entities in the core network, such as P-GW or Policy Charging Rule Function (PCRF)~\cite{ts23.261}.
	
	\section{Conclusions}
	\label{conclusion}
	
	In this paper, we propose the TO algorithm by considering the trade-off between network signaling overhead and femtocell offloading capability. The TO algorithm is evaluated by two performance metrics: signaling overhead reduction ratio $\Theta(t_{o})$ and offloading capability loss ratio $\Lambda(t_{o})$. Our analytical model and simulation results show consistent findings that the proposed TO algorithm is capable of significantly reducing signaling overhead against various UE mobility behaviors while little femtocell offloading capability is compromised, particularly for UEs with more dynamic mobility behaviors (i.e., larger $v_{f}$). Moreover, our work provides guidelines for network operators on how to set an optimal offloading threshold to achieve the best performance.

	\bibliographystyle{IEEEtran}
	\bibliography{ThresholdOffloading_Bibliography}

	\balance
	
\end{document}